\RequirePackage{fix-cm}
\documentclass[smallextended]{svjour3}       
\smartqed  
\usepackage{rotating,graphicx}
\usepackage{subcaption}
\usepackage[title]{appendix}

\usepackage{paralist}
\usepackage{url}
\usepackage{xcolor}
\usepackage{xspace}
\usepackage{graphicx}
\usepackage{url}
\usepackage{algorithm} 
\usepackage{algpseudocode}
\usepackage{amssymb}
\usepackage{booktabs}
\usepackage{array,multirow}
\usepackage{hyperref}

\newcolumntype{L}[1]{>{\raggedright\let\newline\\\arraybackslash\hspace{0pt}}m{#1}}
\newcolumntype{C}[1]{>{\centering\let\newline\\\arraybackslash\hspace{0pt}}m{#1}}
\newcolumntype{R}[1]{>{\raggedleft\let\newline\\\arraybackslash\hspace{0pt}}m{#1}}
\definecolor{mypink1}{rgb}{0.858, 0.188, 0.478}
\definecolor{mypink2}{RGB}{219, 48, 122}
\definecolor{mypink3}{cmyk}{0, 0.7808, 0.4429, 0.1412}
\definecolor{mygray}{gray}{0.6}

\newcommand{\cybersecurity}{\textit{cyber security}\xspace}
\newcommand{\Cybersecurity}{\textit{Cyber security}\xspace}

\newcommand{\ESTABLO}{\texttt{ESTABLO}\xspace}
\newcommand{\RIVALS}{\textit{RIVALS}\xspace}
\newcommand{\SCATTERED}{\textit{DIFFRACTED}\xspace}
\newcommand{\DarkHorse}{\textit{DARK-HORSE}\xspace}
\newcommand{\DARKHORSE}{\DarkHorse\xspace}
\newcommand{\AVAIL}{\textit{AVAIL}\xspace}

\newcommand{\TOPIC}{Adversarial Genetic Programming for Cyber Security\xspace}
\newcommand{\ECS}{\textit{\TOPIC}\xspace}
\newcommand{\paradigm}{\ECS}

\newcommand{\CAs}{Coevolutionary algorithms\xspace}
\newcommand{\ca}{coevolutionary algorithm\xspace}

\newcommand{\CompCA}{Competitive coevolutionary algorithm\xspace}
\newcommand{\compCA}{competitive coevolutionary algorithm\xspace}

\newcommand{\compCAs}{competitive coevolutionary algorithms\xspace}
\newcommand{\TabCompCA}{Comp Coev\xspace}

\newcommand{\CoopCAs}{Cooperative coevolutionary algorithms\xspace}

\newcommand{\vx}{\mathbf{a}}
\newcommand{\vy}{\mathbf{d}}
\newcommand{\objfct}{\mathcal{L}}

\newcommand{\mx}{\mathbf{A}}
\newcommand{\my}{\mathbf{D}}

\begin{document}

\title{\TOPIC:  A Rising Application Domain Where GP Matters
}

\titlerunning{\TOPIC}        

\author{Una-May O'Reilly \and Jamal Toutouh \and Marcos Pertierra \and Daniel Prado Sanchez \and Dennis Garcia \and Anthony Erb Luogo  \and Jonathan Kelly \and Erik Hemberg 
}


\institute{Una-May O'Reilly \at
              MIT CSAIL \\
              Tel.: +617-253-6437\\
              Fax: +123-45-678910\\
              \email{unamay@csail.mit.edu}           
           \and
           Erik Hemberg \at
              MIT CSAIL \\
              Tel.: +123-45-678910\\
              Fax: +123-45-678910\\
              \email{hembergerik@gmail.com}           
}

\date{Received: date / Accepted: date}

\maketitle

\begin{abstract}
Cyber security adversaries and engagements are ubiquitous and ceaseless. We delineate \textit{\TOPIC}, a research topic that, by means of genetic programming (GP), replicates and studies the behavior of cyber adversaries and the dynamics of their engagements.  \TOPIC encompasses extant and immediate research efforts in a vital problem domain, arguably occupying a position at the frontier where GP matters. Additionally, it prompts research questions around evolving complex behavior by expressing different abstractions with GP and opportunities to reconnect to the Machine Learning, Artificial Life, Agent-Based Modeling and Cyber Security communities. We present a framework called \RIVALS which supports the study of network security arms races.  
Its goal is to elucidate the dynamics of cyber networks under attack by computationally modeling and simulating them. 
\end{abstract}

\section{Introduction}
\label{sec:introduction}

\begin{figure}[!b]
  \centering
 \begin{subfigure}[b]{0.41\textwidth}
    \includegraphics[width=\textwidth]{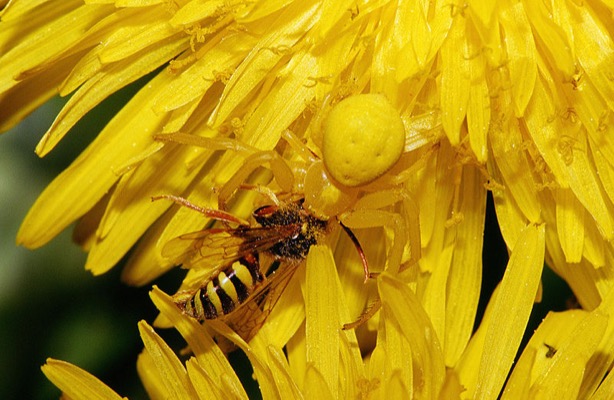}
    \caption{A spider(Misumena vatia)~\cite{spider}}
    \label{fig:spider}
  \end{subfigure}
  \hfill
  \begin{subfigure}[b]{0.41\textwidth}
    \centering
    \includegraphics[width=0.98\textwidth]{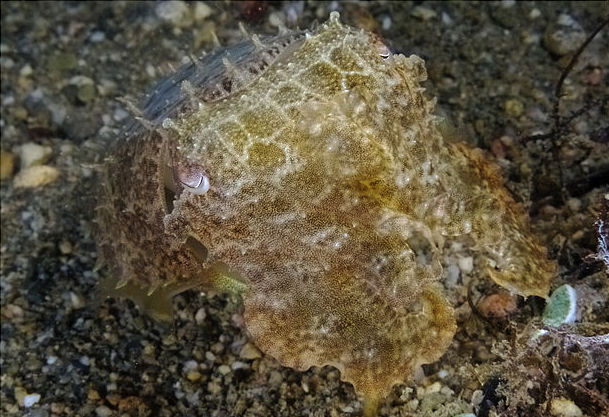}
    \caption{Cuttlefish~\cite{cuttlefish}}
    \label{fig:cuttlefish}
  \end{subfigure}
  \label{fig:animals}
  \begin{subfigure}[b]{0.41\textwidth}
    \includegraphics[width=\textwidth]{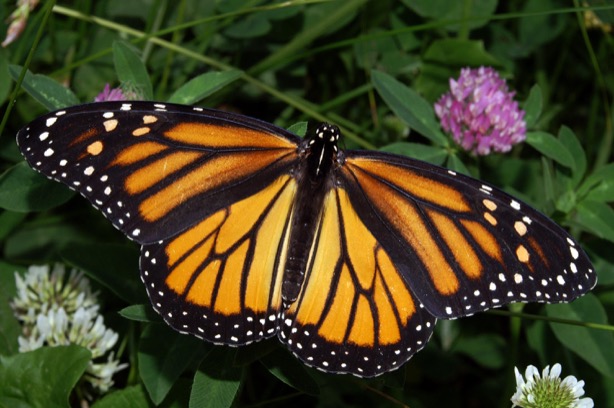}
    \caption{A Monarch butterfly~\cite{monarch}}
    \label{fig:monarch}
  \end{subfigure}
  \hfill
  \begin{subfigure}[b]{0.41\textwidth}
    \centering
    \includegraphics[width=0.86\textwidth]{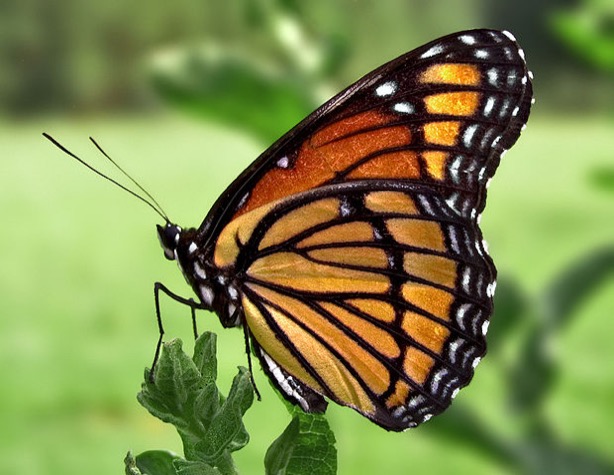}
    \caption{A Viceroy Butterfly~\cite{viceroy}}
    \label{fig:viceroy}
  \end{subfigure}
  
  \begin{subfigure}[t]{0.41\textwidth}
    \includegraphics[width=\textwidth]{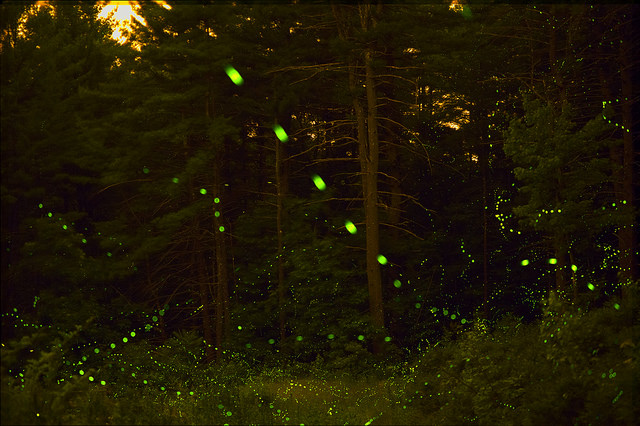}
    \caption{Fireflies~\cite{fireflies}}
    \label{fig:fireflies}
  \end{subfigure}
  \hfill
  \begin{subfigure}[t]{0.41\textwidth}
    \centering
    \includegraphics[width=0.89\textwidth]{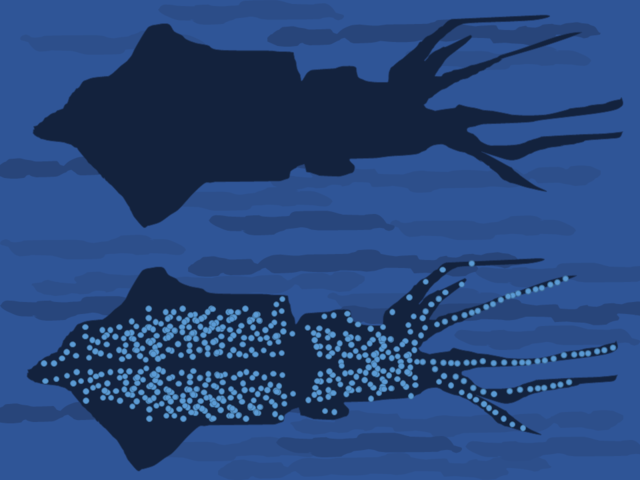}
    \caption{Firefly squid~\cite{firefly_squid}, Appendix~\ref{app:firefly-squid}.} 
    \label{fig:firefly_squid}
  \end{subfigure}
  \label{fig:adaptations}
  \caption{Adversarial defense and attack adaptions. Coloration adaptation -- camouflage (a), (b) and biomimicry (c), (d). Bioluminescence defensive (e), (f).}
\end{figure}

The natural world yields many examples of adversarial advantages arising through evolution. These are testaments to biological arms races that have played out with complex dynamics, typically over macroscopic time scales.  There is a wide range of advantages evolved by predators and prey including examples of animal coloration -- camouflage that is static such as the yellow crab spider (\textit{Misumena vatia}) and dynamic such as the cuttlefish (see Fig.~\ref{fig:spider},\ref{fig:cuttlefish}), bioluminescence (e.g. mid-water crustaceans and fireflies, Fig~\ref{fig:fireflies},\ref{fig:firefly_squid}), and mimicry (e.g. the Monarch and Viceroy butterflies, Fig~\ref{fig:monarch},\ref{fig:viceroy}).

Contemporary human-made ecosystems also yield arms races that arise from adversarial competitions.  In the U.S. automotive industry, firms engage in an arms race to develop innovative new products ahead of the competition.  The relationship between innovations and firm performance has been analyzed from a coevolutionary perspective
\cite{doi:10.1111/jpim.12080} (See Appendix~\ref{app:a}).  
The \textit{circular economy }
and its related \textit{cradle-to-cradle} idea describe manufacturer-customer relationships that cooperatively evolve while business ecosystems within the economy also evolve, though competitively~\cite{Petrlic2016}.
In an example from taxation, tax shelters prey upon unintended lapses of the tax code  until defensive mediations address them. On occasion, however, these mediations either introduce new ambiguity or signal evaders to target a different weakness or inefficiency  and the arms race continues ~\cite{wiki:SoB,wright2013financial}.


One of the contemporary crucibles
of adversarial activity and arms races is cyberspace. It is a complex, human-made ecosystem that spans networks (including the Internet), edge devices, servers, clouds, supercomputers.  Each sub-ecosystem has sets of resources that collectively are a source of contention.   Each includes human actors in circumstances where their software is proxy for their part in cyber adversarial engagements.  Actors can have conflicting objectives.   From a security perspective one population of actors have benign intentions and the other has malicious intentions.  Benign ``cyber actors'' are conforming within social and legal norms. They may control data which they need to keep confidential or private. They may operate networks,  or websites in order to conduct their enterprise's business -- digital commerce. Different devices e.g. cellphones and laptops support everyday social and economic functions. Underpinning all these activities are computational hardware and software. Through attacks on these resources, malicious actors seek to disrupt, infiltrate and exfiltrate to steal, poison, extort so they can profit financially and advance their causes.  

Focusing on network security, one pernicious kind of network attack is named Denial of Service (DOS). The goal of a DOS attack is to consume resources of a target  so the target has none left to serve legitimate clients.  
At the extreme end of the range of volume are Distributed DOS (DDOS) attacks. These are extremely aggressive but more rare, likely due to the cost of setting them up. At a technical level, DDOS attacks are composed from a software toolkit. One set of software comprises multiple tactics that allow  a network of compromised servers, i.e. a botnet,  to be stealthily established. Another has tactics that enable, from a command and control location, a (human) controller to direct the bots to launch DOS attacks. These tactics can even react to defensive measures being deployed during an attack. The controller strategically selects the tactics and aims them at a target.

It is arguable that DDOS attacks are analogous to population members of a species, with tactical variation akin to biological variation. This population arguably undergoes evolutionary adaptation because ineffective attacks are not reused and effective ones undergo adaptation in order to circumvent defensive measures (i.e. reproductive selection and variation are present). With each new generation of attacks, the defensive measures themselves adapt or evolve then they face the next round of adaptation based on the toolkit. The attacker’s general goal is to minimize its use of resources while maximizing its denial of service and the defender’s general goal is to minimize the impact of the attack. In terms of dynamics, an escalating, adversarial arms race around engagements based on the conflicting objectives consequently emerges. 

\begin{figure}[tb]
\centering
\includegraphics[width=0.75\textwidth, trim={0cm 0cm 0cm 1.1cm}, clip]{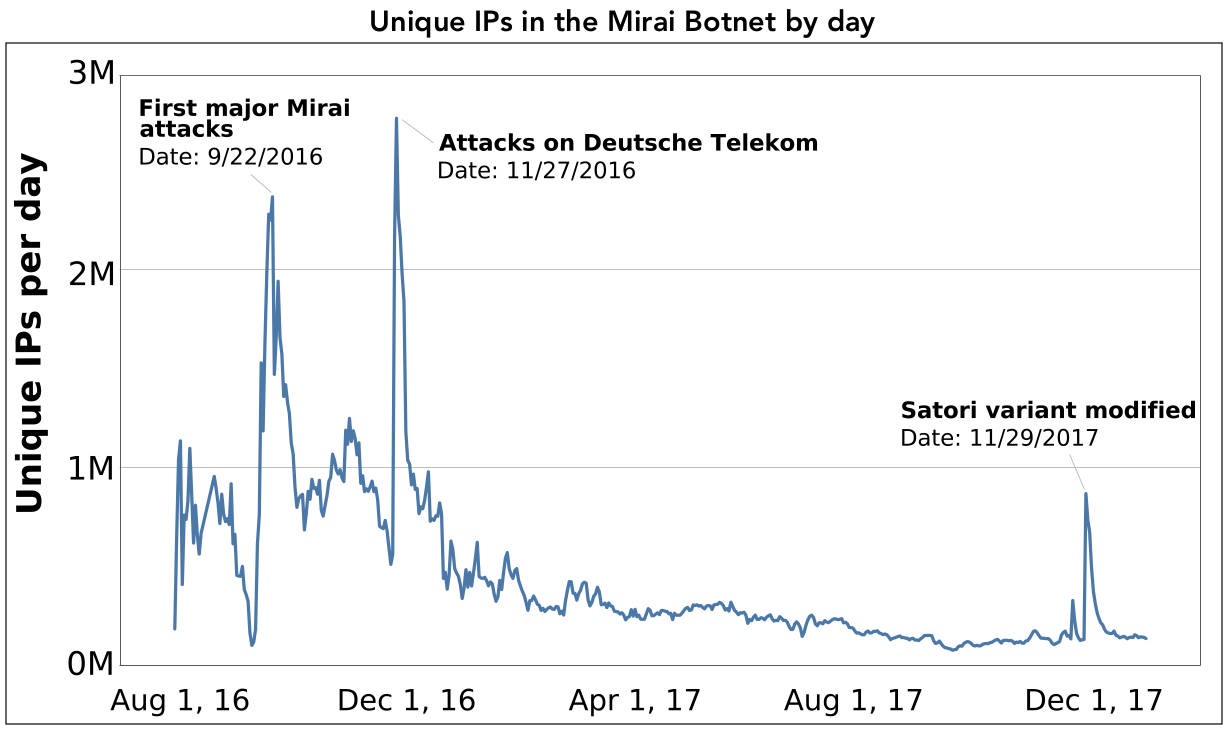}
\caption{Time series of size of bots (number of unique IPs of the compromised servers) in the MIRAI net.  The size escalates prior to an attack. It plummets as the attack is repelled but adaptations allow it to sustain its size then regrow~\cite{AKAMAIreport}, modified for legibility~\cite{akamai2017_Q1}.}
\label{fig:DDOSTimeSeries}
\end{figure}

\begin{figure}[htbp]
	\begin{center}
		\includegraphics[width=0.65\textwidth]{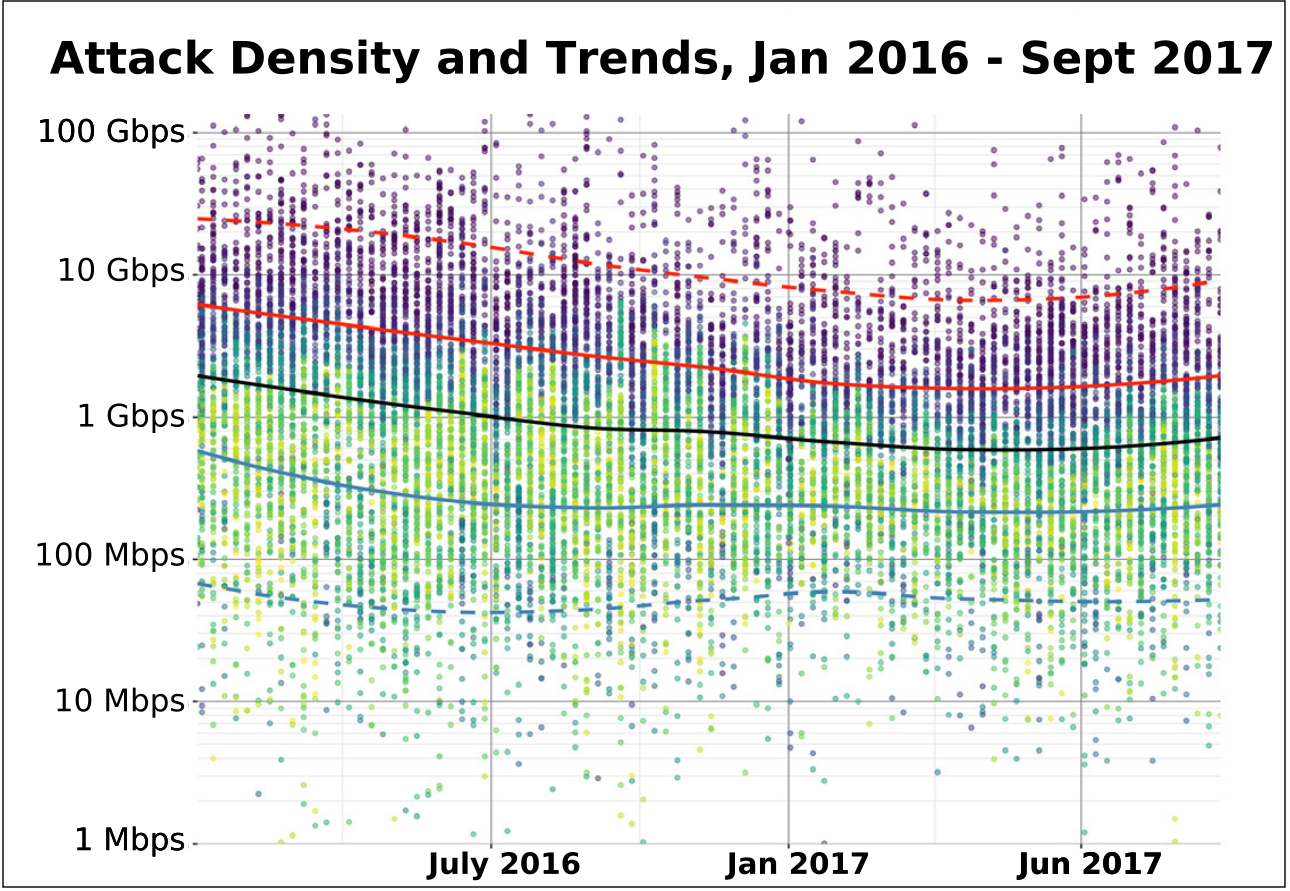}
		\caption{Evolution is churning underneath the surface where we see DDOS attack densities over time, modified for legibility~\cite{akamai2017_Q3}.}
		\label{fig:akamai-ddos-trends}
	\end{center}
\end{figure}

There is reliable and thorough documented evidence of DDOS attacks and counter measures that support an evolutionary interpretation, at a high level of abstraction~\cite{antonakakis2017understanding}.  MIRAI~\cite{url:mirai} is a notorious DDOS, receiving publicity for two high profile attacks: Krebs~\cite{url:krebs} and DYN~\cite{url:dyn}.  From its inception to the present, MIRAI has adapted its tactics and targets to thwart defensive counter-measures, see Fig.~\ref{fig:DDOSTimeSeries}.   Fig.~\ref{fig:akamai-ddos-trends} shows a Red Queen like DDOS dynamics -- while evolution is churning beneath the surface, the number of attacks and their diversity of volume barely changes. Defensive gains are relative and transient. 

\sloppy DOS attacks sit amongst an alarming array of other cyber attack types, such as Advanced Persistent Threats (APT) and Ransomware (see Appendix~\ref{app:apts-ransmoware}) that exhibit intelligent and agile attacks as well as attack versus defense arms race dynamics. 
Across the entire cyber ecosystem there is ceaseless evolution of kinds, arguably species, of attacks and co-evolving responses. 
Beyond their inconvenience, attacks risk lives, have financial costs, threaten businesses, impinge upon privacy and disrupt legitimate activity.

One way to improve defenses requires understanding better how an attacker behaves and how competing attacks and defenses react to one another. Feeding back information on \textit{what could happen} is extremely valuable. It elucidates potential or past behaviors. For defensive design, knowing what could happen can help by \textit{identifying unforeseen or particularly critical attacks}. Knowledge of arms race dynamics can inform the \textit{adversarially hardening} of defensive designs \textit{ahead of time}.  

Understanding what could happen requires examining, actually playing out, or simulating, the dynamics of ongoing engagements between multiple parties with varying behaviors. Genetic programming (GP) can meet these requirements. It is ideally suited to represent the behavior of an adversary when it engages an opponent. Within a competitive evolutionary algorithm, GP representations can serve as the units of its selection and variation and GP operators can serve as the means of generating new solutions from existing ones. 

Cyber security needs a way of exploring and executing attack versus defense engagements. GP is a paradigm that handles the performance selection and variation of units that \textbf{behave}! It can explore by directly manipulating executable functions without syntactic destruction or explicit program knowledge. It can express behavioral solutions drawn from a search space of hierarchical, varying length, executable structures  and it has genetic operators that are able to blindly, at the representation level, generate novel syntactically correct solutions. Cyber security needs an expressively powerful way to describe abstracted attacks and defenses in a way that they are also able to actually compete against each other. 
GP can fulfill this because it can accommodate any abstraction that can be described as a set of  \textit{functions} and \textit{terminals} or a computational language expressed by a grammar. 

\begin{figure}[tb]
\centering      
\includegraphics[width=0.6\textwidth, trim={5.1cm 6.6cm 13cm 3.1cm},clip]{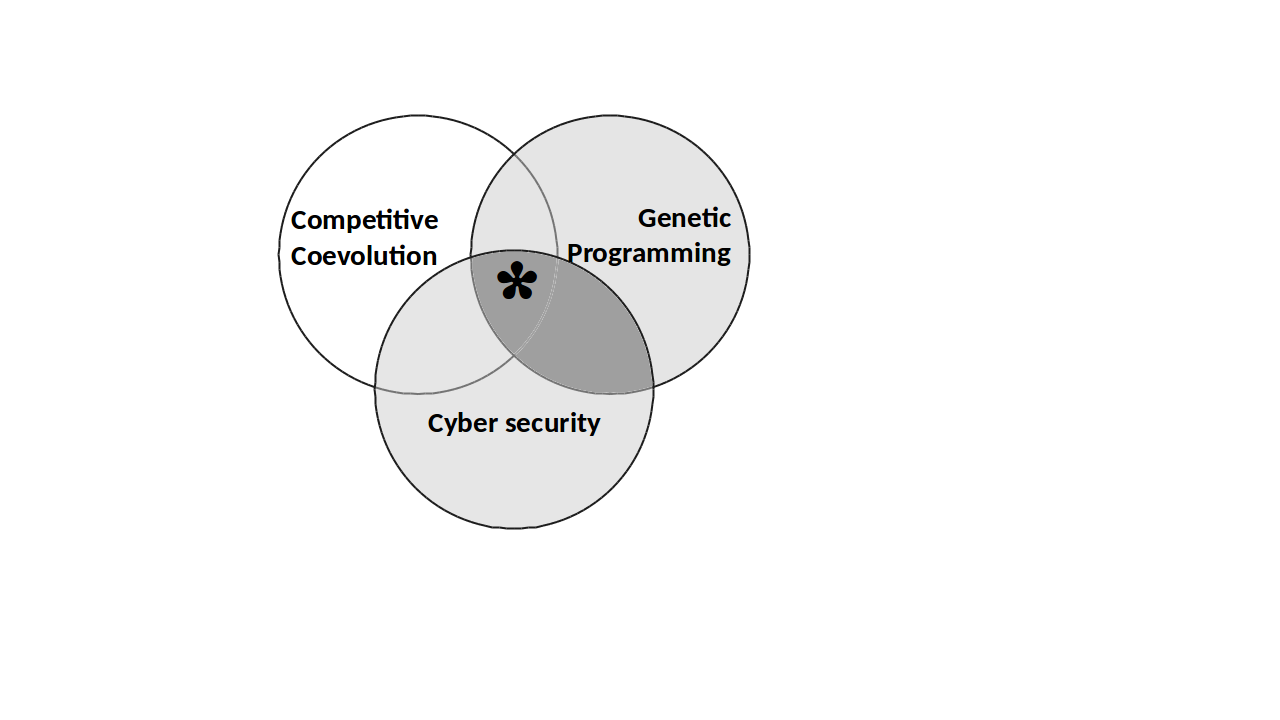}
\caption{Domain intersections of \textit{\TOPIC}.}
\vspace{-0.3cm}
\label{fig:venndiagram}
\end{figure}

Combinations of adversarial evolutionary algorithms and GP matched with the rich problem domain of cyber security thus meld into an increasingly critical intersection with an agenda of compelling and still untapped scope, see Fig.~\ref{fig:venndiagram}. We name this topic \textit{\TOPIC}. It is \textit{not} a new topic; it is the delineation of extant and current research approaches. We call attention to them because cyber security is urgent and escalating in challenge, and despite contributions to date, many novel approaches are required.  At the 20 year mark of GP, when it is no longer necessary to show that GP works, it is timely to direct mature and nascent GP methods plus GP researchers in the direction of this critical domain with real world problems. 
Can  search-based software engineering techniques  gain traction on security exploits by considering adaptation at actual code level? Can GP enable the exploration of  cyber space actors' goals and resulting plans, thereby addressing their intentions and cognitive reasoning?  These questions and ones in between represent an opportunity for GP research that matters.  

\textit{Paper Roadmap} Having motivated \textit{\TOPIC}, in Section~\ref{sec:background},  we first succinctly cover GP and evolutionary algorithms (EAs). To consider adversarial evolution, we then describe competitive coevolutionary algorithms to remind readers of their complexity. We end the section by showing how GP and a coevolutionary algorithm can be combined. In Section~\ref{sec:past}  we survey prior evolutionary computation research into Artificial Intelligence (AI) and games (relevant because games are also competitive contexts), and software engineering (related because tests have been evolved to compete with program solutions to spur correctness).  We then survey research around and within our central topic with Table~\ref{tab:past} summarizing. 
Sections~\ref{sec:background} and~\ref{sec:past} inform the design motivations underlying a framework we present in Section~\ref{gp_coev_sec:past}. Named \RIVALS, it combines competitive coevolutionary algorithms and GP for network security design. We describe three systems drawing on the framework and a component that extracts useful solutions to support the design decisions of a network defense designer. We draw to a close with a summary and a broad discussion of possible paths forward within this exciting, important paradigm.

\section{Basis Algorithms for Adversarial Evolution}
\label{sec:background}
Variable length genotypes, such as the hierarchical tree structures introduced by Koza~\cite{koza1992genetic}, define large search spaces and improve the flexibility and power of Evolutionary Algorithms conducted by executable structure search~\cite{Kinnear1999,o1997introduction}. GP is arguably defined by its distinctive structures that are variable length and executable plus its operators that preserve syntactic correctness while changing genotype size and structure. 
An Evolutionary Algorithm~(EA) can evolve individual solutions in the
form of executable structures, fixed length genotypes such as the bit strings used by Genetic Algorithms (GAs)~\cite{Goldberg1989} are inflexible.

Biological coevolution refers to the influences two or more
species exert on each other's evolution~\cite{rosin1997new}.  A seminal
paper on reciprocal relationships between insects and plants coined the
word ``coevolution''~\cite{ehrlich1964butterflies}.  Coevolution can
occur in the context of cooperation, where the species experience mutual
benefit and adaptation, or in the context of competition, where the species negatively interact due to constrained resources that are under shared contention or a predator-prey relationship.

EAs usually abstract the evolutionary process while ignoring coevolution. To evaluate the quality of an individual, they apply an \textit{a-priori} defined fitness function and this function does not reflect the possible interactions between another individual or  a dynamic environment.  Optimality can be formulated by an absolute rather than relative objective. Coevolutionary algorithms extend EAs by basing the fitness of an individual on its interactions with other individuals or the environment.   They mimic the coupled species-to-species interaction mechanisms of natural coevolution in order to solve a niche of search, optimization, design, and modeling  problems~\cite{Antonio2018,popovici2012,rosin1997new,sims1994evolving}. 

Similar to the biology, \ca{s} are categorized as cooperative or competitive.  \CoopCAs abstractly model the beneficial interaction between populations or in environments with time dependent factors~\cite{krawiec2016solving}. They are frequently set up with multiple populations each solving a distinct sub-problem of a larger problem.  We continue this section by describing \compCAs in more detail as it is central to adversarial behavior.  

\subsection{Competitive Coevolutionary Algorithms}\label{sec:coev-algor}

A basic \compCA evolves two coupled populations, each with conventional selection and representation-aligned variation (crossover and
mutation) operators which suit the representation of the population's genotype. One population comprises what is commonly called \textit{tests} and the other \textit{solutions}. We refer to individuals in a \compCA generally as \textit{adversaries}.  In each generation, different competitions are formed by pairing up a test and a solution drawn from their respective populations. This couples the two population as they share a fitness evaluation component.  We shall shortly describe different ways in which these competitions can be set up.    

\begin{figure}[h!]
\begin{subfigure}[b]{0.19\linewidth}
\centering
\includegraphics[width=\textwidth]{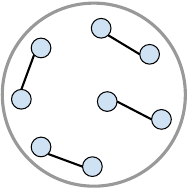} \\
\caption{One-vs-one, for 1 population.}
\label{fig:1pop-1to1}
\end{subfigure}
\hfill
\begin{subfigure}[b]{0.19\linewidth}
\centering
\includegraphics[width=\textwidth]{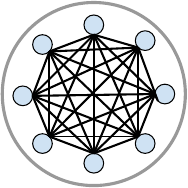} \\
\caption{All-vs-all, for 1 population.}
\label{fig:1pop-alltoall}
\end{subfigure}
\hfill
\begin{subfigure}[b]{0.19\linewidth}
\centering
\includegraphics[width=\textwidth]{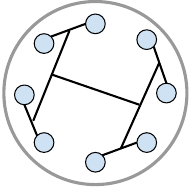} \\
\caption{Tournament, for 1 population.}
\label{fig:1pop-tournament}
\end{subfigure}
\begin{subfigure}[b]{0.19\linewidth}
\centering
\includegraphics[width=\textwidth]{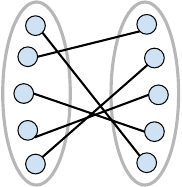} \\
\caption{One-vs-one, for 2 populations.}
\label{fig:npop-1to1}
\end{subfigure}
\hfill
\begin{subfigure}[b]{0.19\linewidth}
\centering
\includegraphics[width=\textwidth]{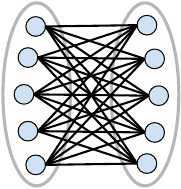} \\
\caption{All-vs-all, for 2 popualtions.}
\label{fig:npop-alltoall}
\end{subfigure}
\hfill
\caption{Competition structures between adversaries/competitors.}
\label{fig:pop-competition-pattern}
\end{figure}

A test competes to demonstrate the solution as incorrect; this is its objective.  The solution competes to solve the test correctly; this is its objective. In a security setting, it may be more apt to translate from \textit{test} and \textit{solution} to \textit{attack} and \textit{defense} as well as to refer to \textit{engagements} rather than \textit{competitions}.   Fitness is  calculated over all of an adversary's engagements. The dynamic of the algorithm, driven by conflicting objectives and guided by performance-based selection and random variation can gradually produce better and more robust
solutions (i.e defenses)~\cite{rosin1997new,sims1994evolving}.  Generally, \compCA{s} suit domains in which there is no exogenous objective measure of performance but wherein performance is relative to others. These have been called \textit{interactive} domains in~\cite{popovici2012} and include games and software engineering.  In most domains a competition is often computationally expensive because it involves simulation or a complex testbed.   We next describe how  competition structuring addresses this challenge.
\paragraph{Competition Structures}
For efficiency, the algorithm designer tries to minimize the number of competitions per generation while maximizing the accuracy of its fitness estimate of each adversary.  The designer is able to control how competitions are structured and how many competitions are used to estimate the fitness of an adversary.  Assuming one or both populations are of size $N$, two extreme structures are: \textit{one-vs-one}, each adversary competes only once against a member of the opposing population, see Fig.~\ref{fig:1pop-1to1} and~\ref{fig:npop-1to1}, and \textit{all-vs-all}, each adversary
competes against all members of the opposing population, see
Fig.~\ref{fig:1pop-alltoall}
and~\ref{fig:npop-alltoall}. \textit{One-vs-one} has minimal fitness
evaluations\footnote{computational cost is shown for two populations.}
($O(N)$) and strong competition bias.  In
contrast, \textit{all-vs-all} has a quadratic number of fitness evaluations, yielding a high
computational cost, $O(N^2)$ but weaker competition
bias~\cite{sims1994evolving}. Other structures provide intermediate
trade-offs between computation costs and competition bias,
e.g. a \textit{tournament} structure ranks individuals based on different
rounds of peer competitions, see Fig.~\ref{fig:1pop-tournament}.

In \cite{mitchell2006coevolutionary}, adversaries termed hosts and parasites
are placed on a \textit{M}$\times$\textit{M} grid with a fixed neighborhood
(size $c$) and one host and parasite per cell. The structure of the competitions is competition among all competitors in the neighborhood.  Fitness evaluations are
reduced to $O(Mc^2)$ by this. An adversary has an outcome for each competition. We next discuss how these outcomes can be united to assign it a fitness score.

\paragraph{Fitness Assignment}
At each generation an adversary needs to be assigned a fitness score derived from some function of its performance outcomes in its competitions. For example, the function can average the performance outcomes or use the maximum, minimum, or median outcome~\cite{axelrod1984evolution}. Other approaches have been defined
depending on the specific problem domain,
e.g.~\cite{fogel2001blondie24,Angeline1993competitive,Cardona2013,lim2010evolving,luke1998genetic,Togelius2007}. A more formal approach, using the \textit{solution} and \textit{test} perspective, describes fitness assignments as \textit{solution
concepts}~\cite{popovici2012}. Solution concepts include:
\begin{inparadesc}
\item best worst case,
\item maximization of expected utility,
\item Nash equilibrium, and
\item Pareto optimality.
\end{inparadesc}

\CAs are more  complex and challenging to direct toward some expected outcome than a single population EA. We next explain the pathologies they exhibit and describe accepted remedies for them.

\paragraph{Pathologies and Remedies}
The interactive aspect of fitness implies that a \compCA lacks an \textit{exact} fitness measurement. While EAs use a  fitness function that allows an individual to be
compared and globally ranked, in \ca{s} two members of the same population, during selection, may be imprecisely compared when they did not compete against the same opponents. Adding more imprecision, regardless of the function that computes a fitness score, an adversary's score may change if it later competes against different opponents.  These properties imply that any ranking of individuals is a noisy estimate. This estimation can lead to \compCAs \textit{pathologies} that
limit the effectiveness of their arms race modeling~\cite{popovici2012}. Furthermore, pathologies can arise from competitive imbalance, particularly when the behavior space of one population is not the same as that of the other. This asymmetry is considered and addressed in \cite{olsson2001co}.
Pathologies include: 
\begin{inparadesc}
\item \textit{disengagement --} occurring when opponents are not challenging enough, eliminating an incentive to adapt, 
or when opponents are too challenging and progress becomes
impossible~\cite{cartlidge2004combating},
\item \textit{cyclic dynamics --} generally appearing in transitive domains (A beats B, B beats C and C beats A) as oscillation of the evaluation metric~\cite{hornby1999diffuse},
\item \textit{focusing} or \textit{overspecialization --} arising when an adversary evolves to beat only some of its opponents, while neglecting the others~\cite{bucci2007emergent}, and 
\item \textit{coevolutionary forgetting  --} occurs when an adversary's ability to beat an opponent evolves away~\cite{boyd1989mistakes}.
\end{inparadesc}

In order to remediate these pathologies, assorted methods have
been proposed~\cite{bongard2005nonlinear,boyd1989mistakes,bucci2007emergent,cartlidge2004combating,ficici2004solution,hornby1999diffuse,krawiec2016solving,williams2005investigating}.
These center on archives or memory mechanisms, maintaining
suboptimal individuals in the population, using a neighborhood
spatial structure, and/or monitoring the progress of the algorithm.

With the basics of GP and \compCA{s} introduced, we are now able to present a combined competitive coevolutionary and GP algorithm.


\subsection{GP and Adversarial Evolution}
\label{sec:gp-advers-evol}
Combining GP and \compCAs enables \cybersecurity arms races to be replicated by
evolving executable adversaries. An example of an alternating
GP \compCA~\cite{arcuri2014co} is shown in
Algorithm~\ref{alg:coevolutionary_algorithm_alternating}.  The
adversaries~(attacker/defender) are coupled and evolve in an
alternating manner. First, the adversaries are initialized. Then, at
each generation, a new attacker population, $\mx_t$, is selected,
mutated, recombined and evaluated against the current defense
population, $\my_{t-1}$. Based on the evaluation the attackers are
replaced. Then, control reverts to the defender population so it can evolve to conclude the generation. This algorithm allows more attack evolution than defensive evolution, reflecting what can happen in real cyber systems. A note of caution about this approach in the context of algorithms, rather than reality, is offered by~\cite{milano2019moderate} where it is observed that moderate environmental variation across generations is sufficient to promote the evolution of robust solutions.


\begin{algorithm}[tb]
	\scriptsize
	\caption{Example of alternating GP \CompCA \newline
	  \textbf{Input:}\newline
          $~~T$: number of generations $~~\objfct$: Fitness function, $~~\mathcal{F}$: Functions, $~~\mathcal{T}$: Terminals\newline
	  $~~\mu$: Mutation probability, $~~\xi$: Crossover probability,~$~~N$: Population size
	}
	\label{alg:coevolutionary_algorithm_alternating}
	\begin{algorithmic}[1]
		\State $\mx_0 \gets [\vx_{1,0}, \ldots, \vx_{N,0}] \sim \mathcal{U}(\{\mathcal{F},\mathcal{T}\})$\Comment{Initialize minimizer(attacker) population}
		\State $\my_0 \gets [\vy_{1,0}, \ldots, \vy_{N,0}] \sim \mathcal{U}(\{\mathcal{F}, \mathcal{T}\})$\Comment{Initialize maximizer(defender) population}
        \State $t\gets 0$ \Comment{Initialize iteration counter}
        \Repeat
		\State $t \gets t + 1$ \Comment{Increase counter}
		\State $\mx_t \gets select(\mx_{t-1}))$ \Comment{Selection}
		\State $\mx_t \gets mutate(\mx_t, \mu))$ \Comment{Mutation}
      		\State $\mx_t \gets crossover(\mx_t, \xi))$ \Comment{Crossover}  
                
		\State $\vx'_*, \vy'_* \gets \arg\min_{\vx \in \mx_t} \arg\max_{\vy \in \my_{t-1}} \objfct(\vx, \vy)$ \Comment{Best minimizer}
                
		\If {$\objfct(\vx'_*, \vy'_*) < \objfct(\vx_{N, t-1}, \vy_{N, t-1})$} \Comment{Replace worst minimizer}
		\State $\vx_{N,t-1}\gets \vx'_*$ \Comment{Update population}
		\EndIf
		\State $\mx_t \gets \mx_{t-1}$ \Comment{Copy population}
		\State $t \gets t + 1$ \Comment{Increase counter before alternating to maximizer}
		\State $\my_t \gets select(\my_{t-1}))$ \Comment{Selection}
		\State $\my_t \gets mutate(\my_t, \mu))$ \Comment{Mutation}
          	\State $\my_t \gets crossover(\my_t, \xi))$ \Comment{Crossover}
                
		\State $\vx'_{*}, \vy'_{*} \gets \arg\min_{\vx \in \mx_{t}} \arg\max_{\vy \in \my_t} \objfct(\vx, \vy)$ \Comment{Best maximizer}
                
		\If {$\objfct(\vx'_{*}, \vy'_{*}) > \objfct(\vx_{N,t}, \vy_{N, t-1})$} \Comment{Replace worst maximizer}
		\State $\vy_{N,t-1}\gets \vy'_{*}$  \Comment{Update population}
		\EndIf
		\State $\my_t \gets \my_{t-1}$ \Comment{Copy population}
		\Until $t \geq T$
		\State $\vx_*, \vy_* \gets \arg\min_{\vx \in \mx_T} \arg\max_{\vy \in \my_T} \objfct(\vx, \vy)$ \Comment{Best minimizer}
		\State \Return $\vx_*, \vy_*$
	\end{algorithmic}
\end{algorithm}

There are some open challenges when applying \ECS.  Some of them
are inherent in the use of EAs and adversarial evolution, such as
the \textbf{subjective solution evaluation}, i.e., individuals
interact with different opponents and the fitness is based on these,
so it only provides a relative
ordering~\cite{popovici2012coevolutionary,popovici2015framework,popovici2017bridging}. Other
amplified challenges are the \textbf{evaluation cost}, since the
individual's fitness is calculated based on its interaction with other
individuals which might require high computational
costs~\cite{bari2018selection,kimclustering,liskowski2016non,mitchell2006coevolutionary,liskowski2017online};
Finally, combining GP and adversarial evolution generates complex
models that complicate the \textbf{algorithm operator and parameter
selection}.
  
This section has described algorithms that form the foundation of
computational adversarial evolution. In the next section we describe
prior Evolutionary Computation work that use GP, competitive
coevolutionary algorithms or a combination of them in the domains of AI
and Games, security, and Software Engineering.

\section{Related Work}
\label{sec:past}
\begin{sidewaystable}
\centering
\scriptsize
\setlength{\tabcolsep}{3pt}
\begin{center}
\caption{Some representative related work to \cybersecurity. It includes three different domains: games, security, and software, and three algorithmic solutions: GP (genetic programming), \TabCompCA (Competitive coevolution), and combination of \TabCompCA and GP. The columns relate to properties described in Section~\ref{sec:coev-algor}, algorithm class, number of populations, representation of the individuals, competition structure and how fitness is assigned to an individual.}
\label{tab:past}
\begin{tabular}{@{}lL{2.2cm}L{3.5cm}L{3.8cm}L{2cm}L{3.6cm}@{}}\toprule
Reference & Algorithm & \# of Populations & Representation & Competition Structure & Fitness Scoring \\
\midrule
\multicolumn{6}{c}{\textit{Games} } \\
\midrule
Axelrod, et al.  
\cite{axelrod1984evolution,axelrod1987evolution} & \TabCompCA GA & One & Bit strings & Tournament & Maximum expected utility\\
Crawford-Marks et al. \cite{crawford2004virtual} & \TabCompCA GP & Two: players and smart-balls & Trees & One vs subset & Maximum expected utility \\
Harper 
\cite{harper2014evolving} & \TabCompCA GP & One & Variable integer vector & One vs a subset & Maximum expected utility  \\
Keaveney \& O’Riordan
 \cite{Keaveney2011} & \TabCompCA GP & One & Trees & One vs subset &
Maximum expected utility \\ 

Lim et al.
 \cite{lim2010evolving} & GP & One & Trees & Playing games &  Pareto optimality \\
Luke et al.
 \cite{luke1998genetic} & \TabCompCA GP & One & Trees & One vs one & Maximum expected utility \\
Miles et al. 
 \cite{miles2007co} & \TabCompCA GA & One & Bit strings & One vs one & Maximum expected utility\\
\midrule
\multicolumn{6}{c}{\textit{Security} } \\
\midrule
Garcia et al. 
\cite{Garcia2017_GECCO} & \TabCompCA GP & Two: Defender \& Attacker & Integers \& BNF Grammar & NA & Multiple comparison \\
Hemberg et al. 
\cite{hemberg2018adversarial} & \TabCompCA GP & Two: Defender \& Attacker & Integers \& BNF Grammar & All vs All & Pareto optimality\\
 Hingston \& Preuss
 \cite{5949747} & \TabCompCA GA & Four: two of learners and two of tests & Pairs of real numbers & One vs all & Best-worst case  \\
 Kewley \& Embrechts
 \cite{1039200} & \TabCompCA GA & Eight: four friendly and four enemy forces & Vector of four parameters & One vs a subset & Maximum expected utility \\
McDonald \& Upton
\cite{mcdonald2005investigating} & \TabCompCA GA & Two: red and blue teams & Vector of parameters & One vs all & Force Exchange Ratio \\
Ostaszewski et al.
  \cite{ostaszewski2007coevolutionary} & \TabCompCA AIS & Two: detectors and anomalies & Vector of parameters & One vs subset & Maximum expected utility   \\

Rush et al. 
\cite{rush2015coevolutionary} & \TabCompCA EA & Two: defenders and attackers & \textit{Not specified} & One vs subset & Maximum expected utility \\ 
 Service \& Tauritz
 \cite{service2009increasing} & \TabCompCA GA  & Two: system hardenings and system faults  & Bit strings: unified power flow controller installation & One vs a subset & Nash equilibrium\\
Suarez-Tangil et al.
\cite{suarez2009automatic} & GP  & One & Trees: intrusion detection rules & Risk assessment & Maximum expected utility  \\


 Winterrose \& Carter
 \cite{winterrose2014strategic} & GA & One: strategies & Bit strings: finite states machine & Fixed scenarios & Success against the defender \\
\midrule
\multicolumn{6}{c}{\textit{Software engineering} } \\
\midrule
Arcuri \& Yao
\cite{arcuri2007coevolving}& \TabCompCA GP & Two: programs and tests & Trees & One vs all & Nash equilibrium \\
Adamopoulos et al. 
\cite{adamopoulos2004overcome} & \TabCompCA GA & Two: programs and tests & Sequences of mutant programs and tests & One vs all & Maximum expected utility \\
Oliveira et al. 
\cite{Oliveira2013} & \TabCompCA GA & Two: programs and tests & Sequences of mutant programs and tests & One vs all & Pareto optimality \\
Wilkerson \& Tauritz
\cite{Wilkerson2010}& \TabCompCA GP & Two: programs and tests &  Trees and lists & One vs a subset & Maximum expected utility \\

\bottomrule
\end{tabular}
\end{center}
\end{sidewaystable}

\begin{figure}[htbp]
	\begin{center}
		\includegraphics[width=0.35\textwidth, trim={7cm 6.2cm 13.8cm 0.1cm}, clip]{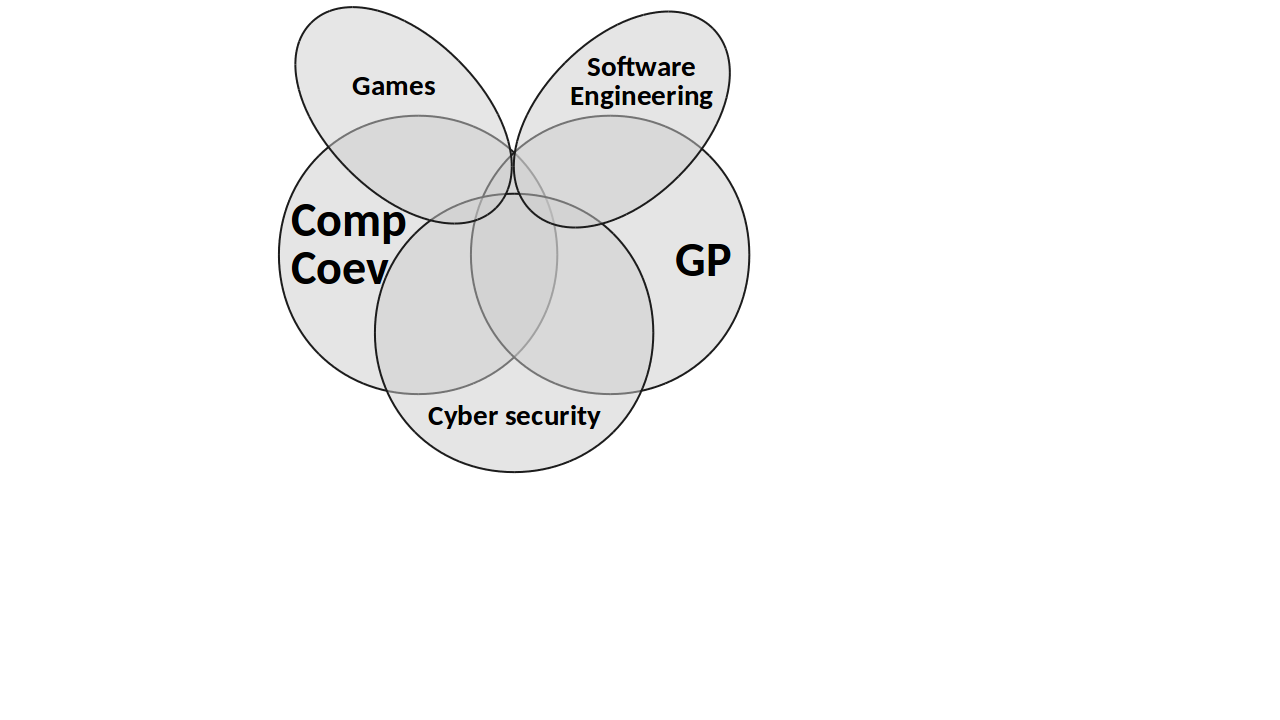} \caption{Venn
			diagram showing the intersections of domains in \ECS related work.} \label{fig:venndiagram-rw}
	\end{center}
\end{figure}

We now turn our attention to prior work relating to the theme of adversarial contexts. We organize by application domain, starting with domains which are competitive in nature but not cyber security: AI and games (\ref{sec:games}) and Software Engineering (\ref{sec:softw-engin-}).  Most examples take technical approaches that use \compCAs or GP.  They provide essential illustrations of how the algorithms can be used as building blocks.  
Table~\ref{tab:past} catalogs the related work by domain, algorithm, representation, competition structure, and fitness scoring.  Fig.~\ref{fig:venndiagram-rw} illustrates how they intersect thematically.  

\subsection{AI and Games}
\label{sec:games}
We find related work in AI and games because many games are competitive and game playing is a fertile research ground
 for investigating adversarial learning. We embrace a broad definition of \textbf{game} -- encompassing board games, e.g.~\cite{Pollack1998}, video game playing, e.g.~\cite{Keaveney2011,sipper2011evolved} and social science games e.g.~\cite{axelrod1984evolution}.

One of the first \compCAs was used to find efficient strategies for the
Iterated Prisoner’s  Dilemma~\cite{axelrod1984evolution,axelrod1987evolution}. This seminal project used a single population and it structured its strategy competition by running a tournament among the population members. A single population is typical for symmetric games, i.e. games where each player has the same set of moves (behavioral repertoire) and objectives.  We find symmetric board game projects that use single population  \compCA{s} to evolve  Tic-Tac-Toe, Othello and Backgammon players.  The Tic-Tac-Toe project represented players with GP ~\cite{Angeline1993competitive} while both Othello and Backgammon projects used temporal difference learning, which is a gradient-based local search  method~\cite{Krawiec2011,Pollack1998,Szubert2009,Szubert2013,smith1994co}.  A subsequent work used symmetric competitive coevolution in games and demonstrated impressive results through self-play~\cite{chellapilla1999evolution}. 
A noteworthy system at the intersection of  \compCA{s}  and GP used a  single population of agents cooperating as a team to compete for the
``RoboCup97'' soccer competition~\cite{luke1998genetic}. Soccer, being symmetric, would allow the system to potentially train against itself. In \cite{sipper2007designing} the authors designed an evolutionary strategizing machine for game playing and other contexts. In~\cite{smith2017coevolving}, the authors coevolved strategies to solve Rubik's Cube where the test population evolved the Rubik's Cube configurations while the learner population evolved GP individuals to solve the Cube configurations. An ablation study demonstrated the contribution of competitive coevolution over random replacement of Cube configurations.

Cyber security adversaries are arguably asymmetric. In asymmetric games where opponents have different objectives and move sets, we find  multi-population \compCA{s}. These include a \compCA with evolutionary
strategies that evolves Pac-Man versus Ghost Team
players~\cite{Cardona2013} and a PushGP system coevolving player versus smart-ball  for the game ``Quidditch''. The latter is noteworthy for its intersection of a \compCA and GP.  

In some asymmetric games one adversary is external to the learning system and quite complex. This drives single or multi-population \ca systems. A \compCA and GP  study compared using a single population to nine populations when evolving controllers for a car
racing game~\cite{Togelius2007}. The multi-population approach generally produced better controllers. Similarly, real time strategy  games (which provide challenging AI
problems~\cite{Lara-Cabrera2013,Nogueira2016}) have applied a
single-population \compCA and GP to develop automated players that use a
progressive refinement planning technique~\cite{Keaveney2011}.  Other
work~\cite{miles2007co} coevolved real time strategy players by representing the AI agents with Influence Maps~\cite{deloura2001game}. A
real-time Neuroevolution of Augmenting Topologies (rtNEAT) approach evolved neural networks in real time while a game was played~\cite{Stanley2005}.
Behavior Trees (BT) were introduced to encode formal system
specifications~\cite{alex2007behavior}, but have also represented game
AI behavior in commercial games~\cite{Nicolau2017}. GP evolved BT
controllers for ``Mario''~\cite{perez2011evolving} and
``DEFCON''~\cite{lim2010evolving} games. Another study used
Grammatical Evolution to generate Java programs to compete in
``Robocode''. The players were represented by
programs~\cite{harper2014evolving} and a spatial and temporal \compCA 
 and GP were used.

\subsection{Software Engineering and Software Testing}
\label{sec:softw-engin-}

Software testing called \textit{fuzzing} arguably started with ~\cite{hanford1970automatic}
which used grammars to generate program inputs for
tests. This transitioned to direct automated random testing. One relevant security example is
~\cite{godefroid2005dart} that searched for bugs in a security
protocol. Later, constraint based automatic test data generation was used to
generate tests to verify if software variants were safe from
malicious manipulation~\cite{demilli1991constraint}. 

For similar purposes, the  search based software engineering community uses evolutionary computation. Coevolution is used in SBSE with GA representations, see e.g. ~\cite{adamopoulos2004overcome,anand2013orchestrated,Oliveira2013}. The community has used \compCAs and GP to coevolve programs and unit tests from their specification, e.g.~\cite{arcuri2007coevolving,Wilkerson2010,arcuri2014co}. Another study,~\cite{barr2015oracle}, used coevolution to 
distinguish correct behavior from incorrect.  


\subsection{Cyber Security}
\label{sec:security}
We now turn to the domain of cyber security.   There are a number of examples where, without any computational modeling or simulation, cyber security dynamics are described as evolutionary or as arms races.  
According to \cite{willard2015understanding},   \cybersecurity is the task of minimizing an \textit{attack surface} over time. The attack surface is the portion
of a system which has vulnerabilities. Attackers
 attempt to influence the system's nominal state and operation by varying their
interactions with the attack surface by stealth and
non-compliance~(see Appendix~\ref{app:cyber-secur-attack}). \Cybersecurity perimeter protection (e.g. a firewall) is a battle fought and lost. Attackers are able to now gain access and defenders must resort to strategies that assume the attacker is present but hidden. One tactic to detect
their presence is deception. For this a ``honeypot'' that entraps an attacker by imitating its target is commonly deployed. This arms race escalates
\textit{ad infinitum} as attackers then anticipate what the defenders
have anticipated (e.g. the honeypots) and so on.  

Retrospective security event reports also document coevolution. For example, ~\cite{gupta2009empirical} is an
empirical study of malware  evolution. Arguments for employing
nature-inspired technologies for \cybersecurity that mention how biological
and ecological systems use information to adapt in an unpredictable
world include~\cite{crandall2009ecology,ford2006predation,iliopoulos2011darwin,mazurczyk2015towards,sagarin2012natural}.

A selected set of cyber security research that takes technical approaches is described in the Security subsection of Table~\ref{tab:past}. This set include works that we would not consider Adversarial Genetic Programming for Cyber Security: one work  uses a genetic algorithm and neither coevolution or GP\cite{winterrose2014strategic} and six others use are adversarial in nature, i.e. they use coevolution, but not GP\cite{5949747,1039200,mcdonald2005investigating,ostaszewski2007coevolutionary,rush2015coevolutionary,service2009increasing}. The remaining three projects fit within the topic of Adversarial GP\cite{suarez2009automatic,Garcia2017_GECCO,hemberg2018adversarial}. 
The research within this set, along with other related work, can also be distinguished by what specific application it addresses in the domain of cyber security:
\begin{asparadesc}
\item [Moving target defense~(MTD)] techniques seek to randomize
components to reduce the likelihood of a successful attack, reduce the
attack lifetime, and diversify systems to limit the
damage~\cite{evans2011effectiveness,6673500}.  A MTD study investigates an adaptable adversarial strategy based on
Prisoners Dilemma in~\cite{winterrose2014strategic}. Strategies are
encoded as binary chromosomes representing finite state machines that
evolve according to GA. The study has one adaptive defender population
in GA and few fixed scenarios. 

\item[Network Defense Investigation] is studied with the coevolutionary agent-based network defense lightweight event system
(CANDLES)~\cite{rush2015coevolutionary} a framework designed to
coevolve attacker and defender agent strategies with a custom,
abstract computer network defense simulation.  The \RIVALS network security framework, elaborated in Section~\ref{gp_coev_sec:past} supports three studies into respectively DDOS, deceptive and isolation defense.

\item [Self-Adapting Cyber Defenses] are for example the Helix self-regenerative
architecture~\cite{le2013moving}. Helix shifts the
attack surface by automatically repairing programs that fail test cases using Software Dynamic Translation,
e.g. Strata~\cite{scott2001strata}. Another example of automated fault
analysis and filter generation within a system is named
``FUZZBUSTER''~\cite{musliner2013meta,musliner2014}. Like Helix, FUZZBUSTER is
designed to automatically identify software vulnerabilities and create
adaptations that protect those vulnerabilities before attackers can
exploit them. Both FUZZBUSTER and Helix use a GP system called
GenProg~\cite{weimer2010automatic} for automatically fixing code. 

\item [Physical Infrastructure defense] is studied in terms of how network components can be made resilient in \cite{service2009increasing}.  

\item [Anomaly detection] attempts to discern network or other activity behavior that is out of the ordinary or that differs from normal. In the auto-immune system computational paradigm, normal has been characterized as ``self'' \cite{forrest1996sense}. Anomaly detection has been studied as a one class learning problem by \cite{ostaszewski2007coevolutionary} who use the artificial immune systems paradigm and coevolution for search.  Attempts to build GP anomaly detectors have mostly assumed labeled data sets and they evolve a classifier that labels outputs as normal or not. They frequently encounter a class imbalance issue. This is explicitly addressed in \cite{song2005training}. Later work adopted an explicitly Pareto archive formulation of competitive coevolution\cite{lemczyk2007training}.

\item [Vulnerabilty Testing]
Vulnerability testing involves testing a system with modifications to known exploits or attack vectors.
Examples of developing mimicry attacks to test for vulnerability are found in \cite{kayacik2009can,kayacik2011can}.  The approach used the alarm signal for coevolution of a GP exploit generator. Moreover, GP was limited to instructions that were in legitimate applications, forcing GP to search for exploits described in terms of instructions used by legitimate applications.  Others have also used GP to coevolve port scans against the 'Snort' intrusion detector with the objective of evolving port scans that demonstrate holes in the IDS. See for example \cite{laroche2009evolving}. Vulnerability testing for malware in mobile applications using coevolution appears in \cite{bronfman2018artificial}.

\item [Malware detection] has seen GP used to evolve novel PDF malware to automatically evade classifiers~\cite{xu2016automatically} and to study malware in the form of return-oriented program evolution~\cite{fraser2017return}.

\item [Intrusion detection] Most examples of intrusion detection are addressed as a multi-class classification problem where activity such as network traffic must be labeled as normal, denial of service, botnet or something else. Multi-class classification has been studied using GP without coevolution. For example, \cite{suarez2009automatic} optimizes intrusion detection rules. Others study botnet detection with GP under streaming data label budgets and class imbalance~\cite{khanchi2018botnet}.  With both intrusion detection and anomaly detection, one challenge is obtaining a dataset that truly reflects the network or any other cyber environment. For example, a widely used dataset known at KDD'99 has been criticized for its artificially generated normal data, which did not encompass the diversity in real normal behavior, see~\cite{lippmann20001999,mahoney2003} for more details. It is also challenging to label and maintain datasets, see~\cite{garcia2014empirical} as an example.  The persistent nature of these challenges provides additional motivation for \RIVALS.

\item [Battle management] has historically noteworthy work by ~\cite{applegate1990architecture} which focuses on security in battle management. It was novel in considering semi-autonomous control of several  intelligent agents; plan adjustment based on developments during plan execution; and the  influence of the presence of an adversary in devising plans.  Similar but contemporary work on a computational
military tactical planning system uses fuzzy sets and \compCA with a battlefield tactics simulator for decision support~\cite{1039200}.
\item [Red teaming] is a technique utilized by the military operational analysis
community to check defensive operational plans by pitting actors posing as attackers (a red team) against the defense (a blue team). This activity has evolved from human teams to teams of programmers overseeing automated attacks tactics and defensive measures. See~\cite{url:darpa} 
and
studies~\cite{abbass2015art,nettles2010president} trying to automate the
exercise to use less manpower, or use it more efficiently with
agent-based models~\cite{beard2011enhancing,mcdonald2005investigating,wood2000red,yuen2015automated,6189363,5949747}.

\end{asparadesc}

\paragraph{The Next Section} Prior work in AI and games, cyber security and software engineering domains is helpful in showing how \compCA{s} and GP can be used in adversarial contexts. It also provides a modest number of examples of GP combined with a  \compCA . We now proceed to present a detailed example of research within \textit{\TOPIC}.  Named \RIVALS, it is a software framework for studying network security. \RIVALS addresses an aspect of  central importance to the 2016 DARPA grand challenge in cyber security named ``The World's First All-Machine Hacking Tournament''\cite{url:darpa}. This aspect that adaptation of both sides of a cyber security battle ground must be anticipated and that eventually posture reconfiguration needs to be fully automated.

\section{RIVALS}
\label{gp_coev_sec:past}

Consistent with the \paradigm paradigm, our goal is to study the dynamics of cyber networks under attack by computationally modeling and simulating them. Ultimately we aim to provide defenders with information that allows them to anticipate attacks and design for resilience and resistance before their attack surfaces are attacked.  We exploit \compCA{s} and GP for these purposes within a framework named \RIVALS~\cite{alec18,pertierrathesis2018,pradothesis2018}.  

\begin{figure}[tb]
  \centering
  \includegraphics[width=0.9\textwidth]{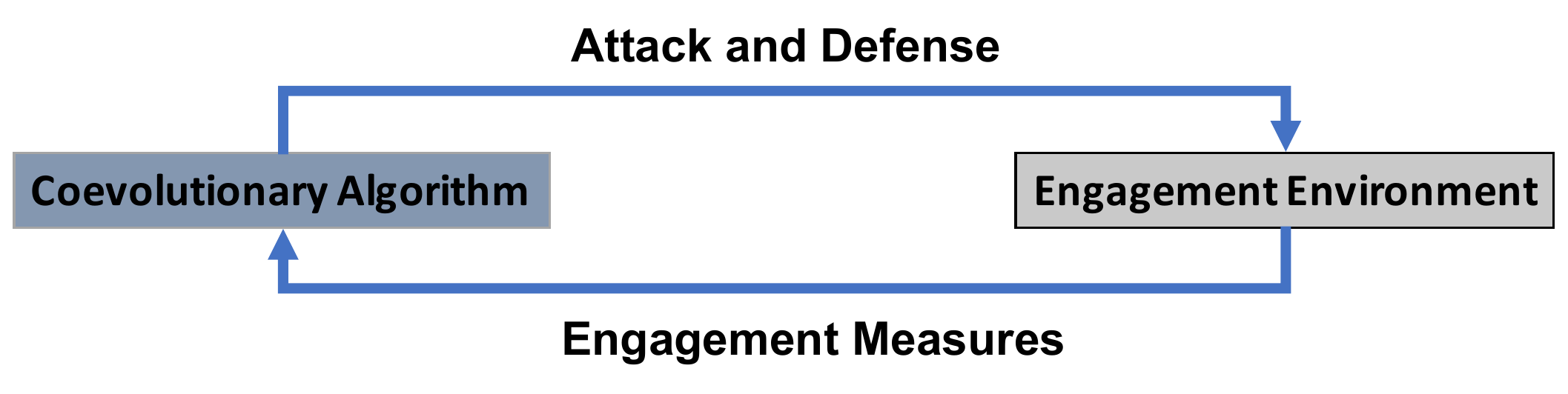}
  \caption{High level view of \RIVALS framework.  The left hand side  component is a competitive coevolutionary algorithm that evolves two competing populations: attacks and defenses. The right hand side component from a computational perspective is a modeling or simulation environment. The environment is initialized with a mission and can be reset each time it is passed an attack-defense pairing to run an engagement. It first installs the defense, then it starts the mission and the attack. It evaluates the performance of the adversaries relative to their objectives and returns appropriate measurements. }
  \label{fig:coevolutionary_framework}
\end{figure}

Conceptually, the framework consists of three elements:
\textit{adversaries},\textit{ engagements and their
environments}, and \textit{competitive coevolution}.  These fit together as two connected modules -- one executing the coevolutionary algorithm and executing the engagements, and the other executing the engagements with the  competitive coevolutionary algorithm directing attack and defense engagement pairings, see Fig.~\ref{fig:coevolutionary_framework},.

We now proceed to describe each of these elements and \RIVALS' use cases.

\subsection{Elements of \RIVALS}

\begin{figure}[tb]
  \centering
  \includegraphics[width=0.9\textwidth]{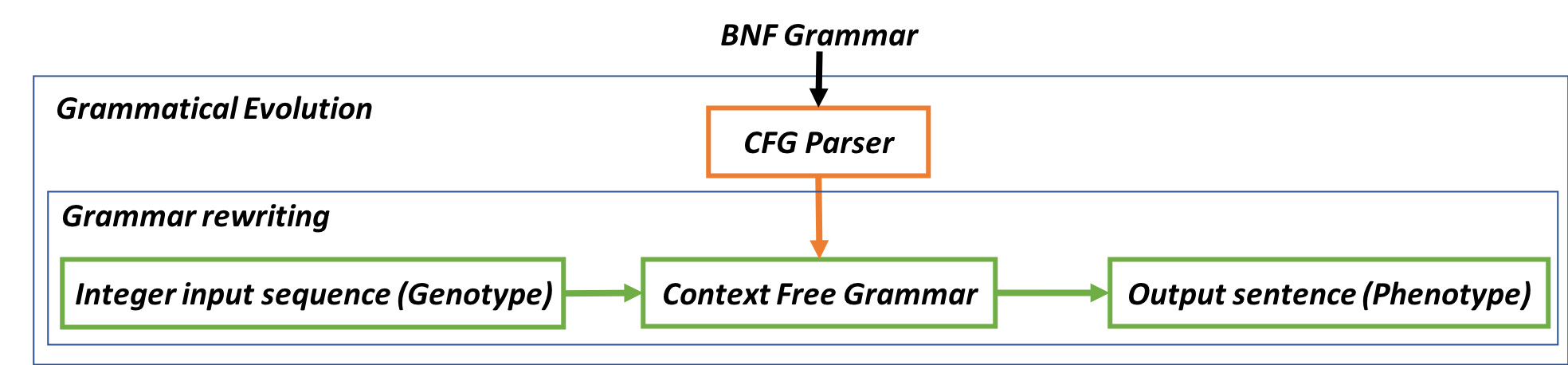}
  \caption{Grammatical evolution string rewriting. A context free grammar rewrites an integer sequence (genotype) to an output sentence (phenotype).}
  \label{fig:GE}
\end{figure}

\paragraph{Adversaries: } The system consists of two adversarial populations -- \textit{attacks} and \textit{defenses}. The primary objective of an \textit{attack} is
to impair a mission by maximizing disruption of some resource. The
primary objective of a \textit{defense} is to complete a mission by
minimizing disruptions. Both attack and defense can have secondary
objectives based on the cost of their behavior.  We design attack and
defense behavior by defining grammars that can express the different
variations of attacks and defenses. An attack or defense is a
``sentence" in the grammar's language. We implement a rewriting process (``generator'')  to
form sentences from the grammar~\cite{GREV}, see Fig.~\ref{fig:GE}. The sentences
(i.e. attacks and defenses) are functions that are executable in the
layers of the engagement environment.

The grammar has Backus Naur Form (BNF) and is parsed to a context free
grammar structure.  The (rewrite) rules of the generator express how a
sentence, i.e. attack or defense, can be composed by rewriting a start
symbol.  The adversaries are represented with variable length integer
vectors. The generator decodes these vectors to control its
rewriting. As a result, different vectors generate different attacks
or defenses.  For different use cases, it is only necessary to change
the BNF grammar, engagement environment and fitness function of the
adversaries.  This modularity, and reusability of the parser and
rewriter are efficient software engineering and problem solving
advantages. The grammar additionally helps communicate to the designer
how the system works and its assumptions, i.e. threat model. This
enables conversations and validation at the domain level with experts
and increases the confidence in solutions from the system.

\paragraph{Engagements and the Engagement Environment}\label{sec:engage}

An engagement is an attack on a defense. Engagements take place in
an \textit{engagement environment} which is initialized with a mission
to complete and a set of resources such as network services that
support the mission.  The defense is first installed in the
environment and then, while the mission runs, the attack is
launched. The scenario (mission and resources) and attacks are then
executed.  Engagements have outcomes
that match up to objectives; they are phrased in terms of mission
completion (primary objective) and resource usage (second
objective). Implementation-wise, the engagement environment component
can support a problem-specific network testbed, simulator or
model. Mod-sim is appropriate when testbeds incur long experimental
cycle times or do not abstract away irrelevant detail.

\paragraph{Coevolutionary Algorithms}\label{sec:coev}

\RIVALS maintains two \textit{populations} of competing attackers and
defenders. It calculates the fitness of each population member (in
both attack and defense populations) by assessing its ability to
successfully engage one or more members from the adversarial
population, given its objective(s). It also directs selection and
variation. \RIVALS~\cite{pradothesis2018,pertierrathesis2018} utilizes
different coevolutionary algorithms to generate diverse behavior. The
algorithms, for further diversity, use different ``solution
concepts'', i.e. measures of adversarial success. However, engagements
are often computationally expensive and have to be pairwise sampled
from two populations at each generation, a number of enhancements enables
efficient use of a fixed budget of computation or
time. 

\paragraph {\RIVALS{'} Compendium}\label{sec:compendium}

Solely emulating or simulating cyber arms races is not sufficient to practically inform the design of better, anticipatory defenses. In fact, competitive coevolution poses general challenges when used for design optimization. The following ones in \RIVALS make it difficult to present a designer with clear information derived solely from multiple simulation runs ~\cite{sanchez2018competitive,pradothesis2018}:
\begin{enumerate}
\item Attacks (and defenses) of different generations are not comparable because fitness is based  solely on the composition of the defense (attack) population at each generation.  So no clear champion emerges from running the algorithm.
\item From multiple runs, with one or more algorithms, it is unclear how to automatically select a ``best'' attack or design.
\end{enumerate}

\begin{figure*}[t]
  \centering
  \includegraphics[width=.99\textwidth]{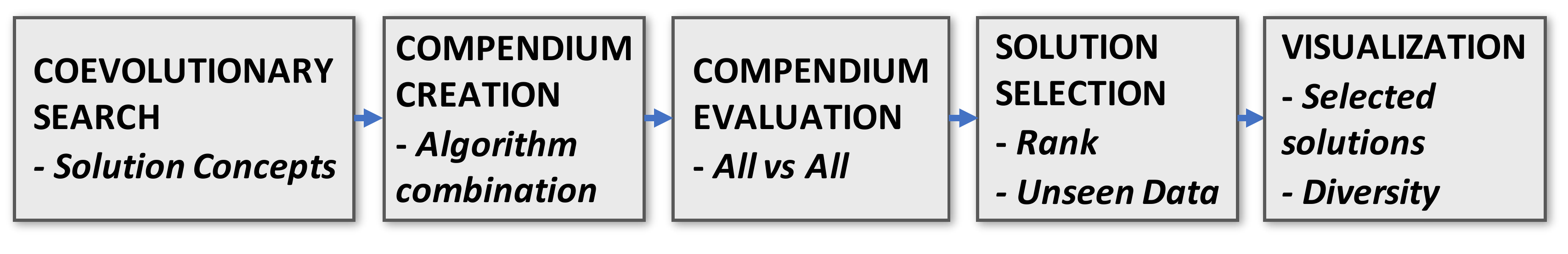}
  \caption{Overview of the \ESTABLO framework used by \RIVALS for decision support
    through selection and visualization by using a compendium of
    solutions from coevolutionary algorithms.}
  \label{fig:ESTABLO}
\end{figure*}

To this end, \RIVALS provides an additional decision support component, named \ESTABLO~\cite{pradothesis2018,sanchez2018competitive}, see Fig.~\ref{fig:ESTABLO}. At the implementation level, the engagements and results of every run of any of the system's coevolutionary algorithms are cached. Later, offline,  \ESTABLO filters these  results and moves a subset to its \textit{compendium}. To prepare for the  decision support analysis, it then competes all the attacks in the compendium against all defenses and ranks them according to multiple criteria, e.g. maximum-expected utility, best-worst case. For the defensive designer, it also provides visualizations and comparisons of adversarial behaviors to inform the design process.


\begin{figure}[!h]
  \centering
  \includegraphics[width=0.99\textwidth]{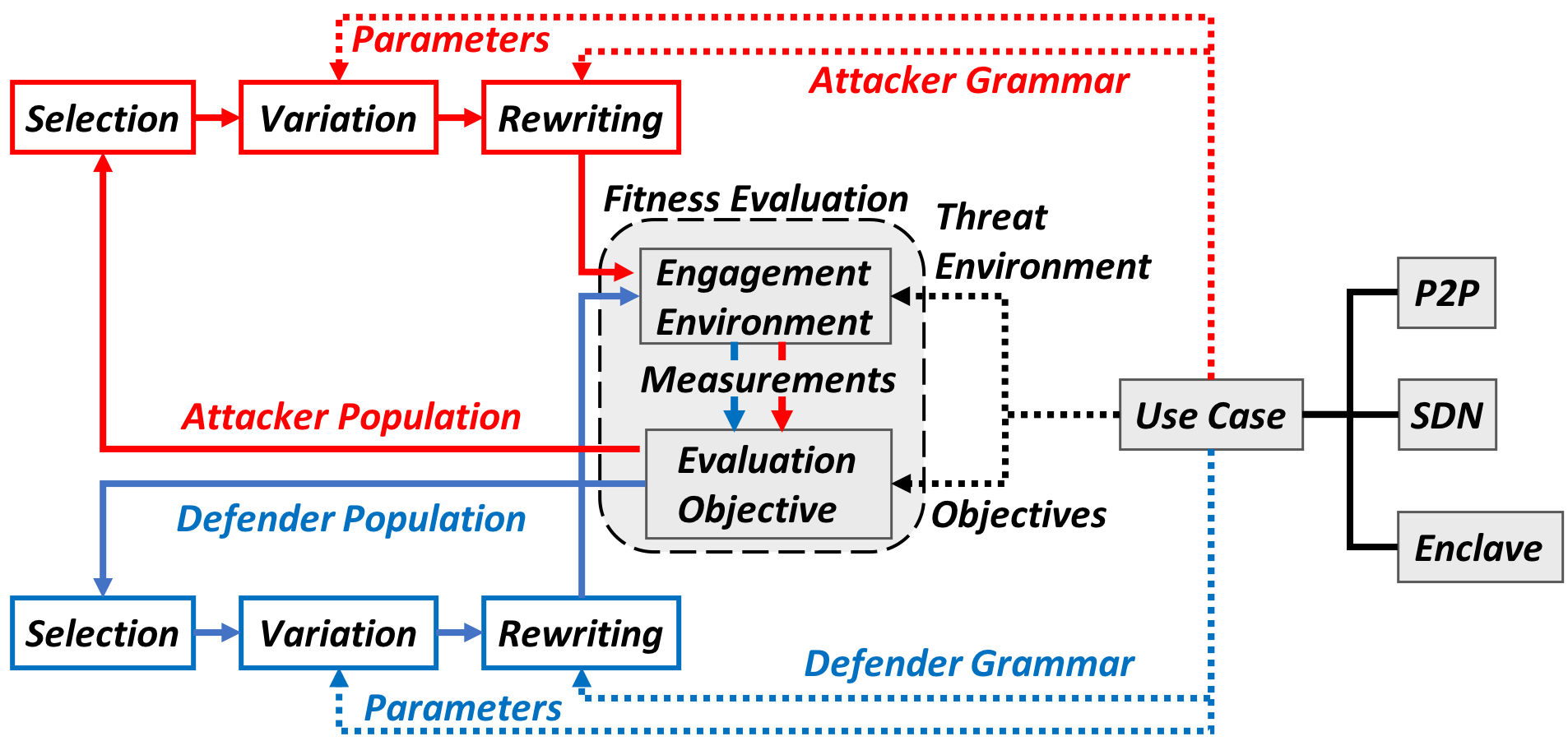}
  \caption{\RIVALS framework component decomposition showing interface with use~cases.}
\label{fig:rivals_concepts}
\end{figure}

\paragraph{Use Cases} \label{sec:usecaseList}
The \RIVALS framework supports use~cases -- investigations into arms races of a particular cyber network context. They each interface with the framework through a set of domain specific information. For each use case, grammars define the behavioral space of the adversaries, objectives define the goals of the adversaries for scoring  fitness  and the threat environment describes its engagement environment.  For each use case, different parameters to control the coevolutionary algorithm are also available.  Fig.~\ref{fig:rivals_concepts}
depicts the high level decomposition of the framework and shows how a use case interfaces with it. Table~\ref{tab:rivals_framework} presents the domain specific information for three use cases. By name, these are:
\begin{itemize}
\item \SCATTERED: Defending a peer-to-peer network against Distributed Denial of
  Service (DDOS) attacks~\cite{Garcia2017_GECCO} (Section~\ref{sec:dos-attacks-peer})
\item \AVAIL: Defenses against device compromise contagion in a segmented
  enterprise network~\cite{hemberg2018adversarial} (Section~\ref{sec:avail-attacks-segm}), and
\item \DARKHORSE: Deceptive defense against the internal reconnaissance of an
  adversary within a software defined
  network~\cite{pertierrathesis2018} (Section~\ref{sec:intern-reconn-softw})
\end{itemize}

The following sections elaborate on these use cases.

\subsection{\SCATTERED -- DOS Attacks on Peer-to-Peer Networks}
\label{sec:dos-attacks-peer}

A peer-to-peer (P2P) network is a robust and resilient means of
securing mission reliability in the face of extreme distributed denial
of service (DDOS) attacks. \SCATTERED~\cite{Garcia2017_GECCO}, assists
in developing P2P network defense strategies against DDOS
attacks. Attack completion and resource cost minimization serve as
attacker objectives.  Mission completion and resource cost
minimization are the reciprocal defender objectives.  DDOS attack
strategies are modeled with a variety of behavioral languages.

A simple language e.g. allows a strategy to select one or
more network servers to disable for some duration. Defenders choose
one of three different network routing protocols: shortest path,
flooding and a peer-to-peer ring overlay to try to maintain their
performance. A more complex language allows a varying number of steps
over which the attack is modulated in duration, strength and
targets. It can even support an attack learning a parameterized
condition that controls how it adapts during an attack,
i.e. ``online'', based on feedback it collects from the network on its
impact.  Defenders have simple languages related to parameterizations
of P2P networks that influence the degree to which resilience is
traded off with service costs.  A more complex language allows the P2P
network to adapt during an attack based on local or global
observations of network conditions.   

\begin{figure}[t]
  \centering
  \includegraphics[width=.99\textwidth]{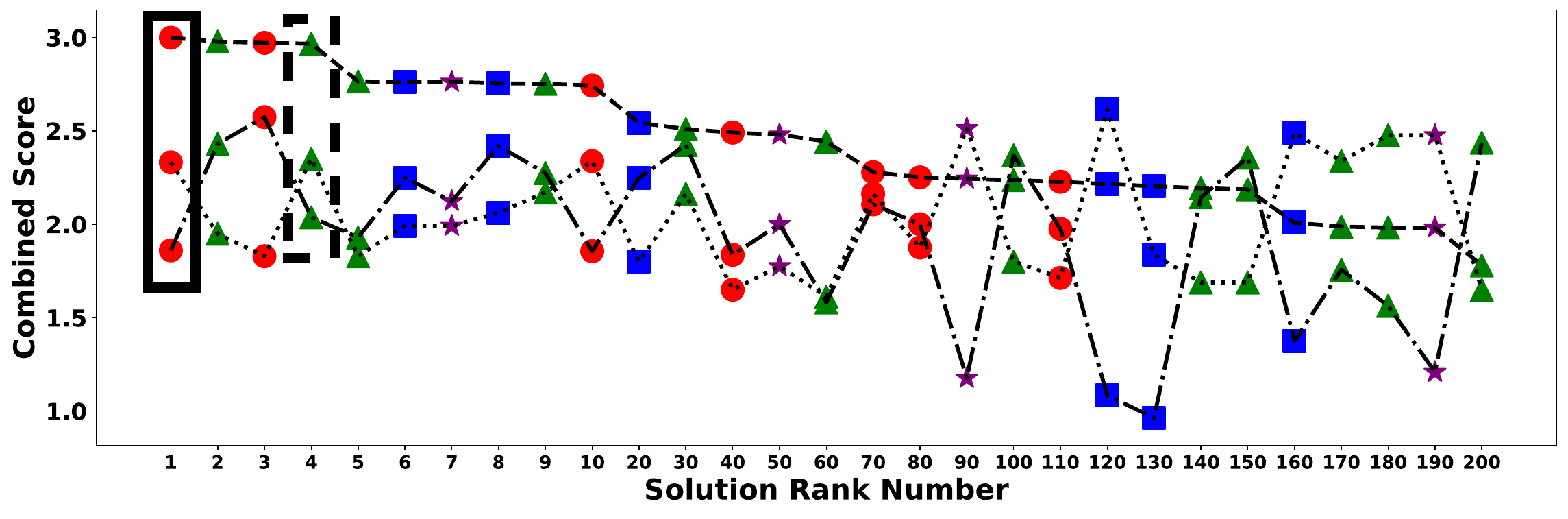}
  \caption{The x axis shows a sorted subsample of attackers
    (note, the top 10 are shown and then every tenth) and
    the y axis shows the ranking score. The ranking is done on the
    scores from the compendium. The values for the same run
    and unseen test sets are shown on separate lines. The algorithm
    used to evolve the attacker is shown by the marker and the
    color. The attacker in the box with the solid line is the top
    ranked solution from the Combined Score ranking schemes. The
    solution in the dashed box is the top ranked solution from the
    Minimum Fitness ranking scheme.}
  \label{fig:compendium_attacker}
\end{figure}

An example of attackers from \ESTABLO on a mobile resource allocation
defense used in \SCATTERED~\cite{sanchez2018competitive} is shown in
Fig.~\ref{fig:compendium_attacker}. The mobile asset placement
defense challenge is to optimize the strategic placement of assets in the
network. While under the threat of node-level DDOS attack, the defense
must enable a set of tasks. It does this by fielding feasible paths
between the nodes that host the assets which support the tasks. A
mobile asset is, for example, mobile personnel or a software
application that can be served by any number of nodes.  A task is, for
example, the connection that allows personnel to use a
software application.

\subsection{\AVAIL - Availability Attacks on Segmented Networks}
\label{sec:avail-attacks-segm}

Attackers often introduce malware into networks. Once an attacker has
compromised a device on a network, they spread to connected devices,
akin to contagion. \AVAIL considers \textit{network segmentation}, a
widely recommended defensive strategy, deployed against the threat of
serial network security attacks that delay the mission of the
network's operator~\cite{hemberg2018adversarial} in the context of
malware spread.

\textit{Network segmentation} divides the network topologically into
\textit{enclaves} that serve as isolation units to deter inter-enclave
contagion. How much network segmentation is helpful is a tradeoff. On
the one hand, a more segmented network provides lower mission
efficiency because of increased overhead in inter-enclave
communication.  On the other hand, smaller enclaves contain compromise
by limiting the spread rate, and their cleansing incurs fewer mission
delays. Adding complexity, given some segmentation, a network operator
can monitor threats and utilize cleansing policies to detect and
dislodge attackers, with the caveat of cost versus efficacy.

\begin{figure}
  \centering \includegraphics[width=0.99\textwidth]{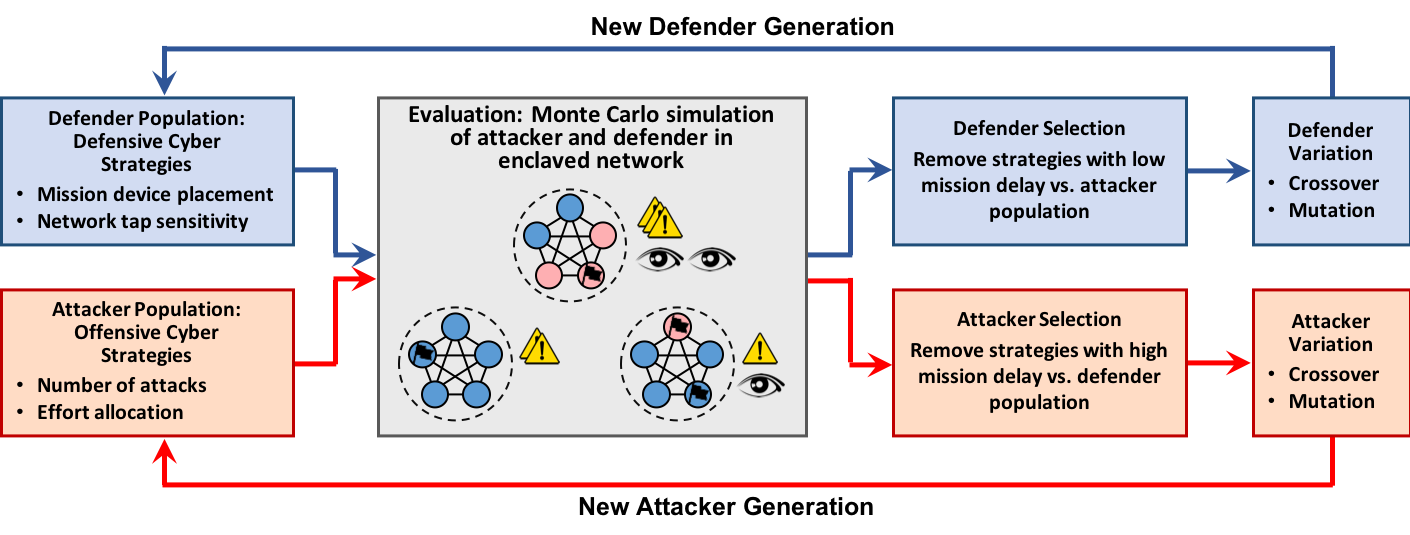} \caption{Flow
  of the \compCA used to evaluate defensive network enclave
  configurations(blue boxes) and contagion attacks(red boxes)
  in \AVAIL~\cite{hemberg2018adversarial}. A Monte Carlo simulation of
  device compromise contagion in a segemented network is used to assign the fitness of
  the adversaries. \AVAIL uses Maximum Expected Utility as a fitness
  score based on the mission delay} \label{fig:avail}
\end{figure}

\AVAIL assumes an enterprise network in carrying out
a \textit{mission}, and that an adversary employs \textit{availability
attacks} against the network to disrupt it. Specifically, the attacker
starts by using an exploit to compromise a vulnerable device on the
network. This inflicts a mission delay when a mission critical device
is infected. The attacker moves laterally to compromise additional
devices to further delay the mission. Fig~\ref{fig:avail}
shows \AVAIL.

\AVAIL employs a Monte Carlo simulation model as its engagement environment.
Malware contagion of a specific spread rate is assumed.  The defender
decides placement of mission devices and tap sensitivities in the
pre-determined enclave segmentation. The attacker decides the
strength, duration and number of attacks in an attack plan targeting
all enclaves.  For a network with a set of four enclave topologies,
the framework is able to generate strong availability attack patterns
that were not identified
\textit{a~priori}. It also identifies effective configurations that
minimize mission delay when facing these attacks.

\subsection{\DarkHorse -- Internal Reconnaissance in Software Defined Networks}
\label{sec:intern-reconn-softw}

\begin{figure}
  \centering
  \includegraphics[width=0.69\textwidth]{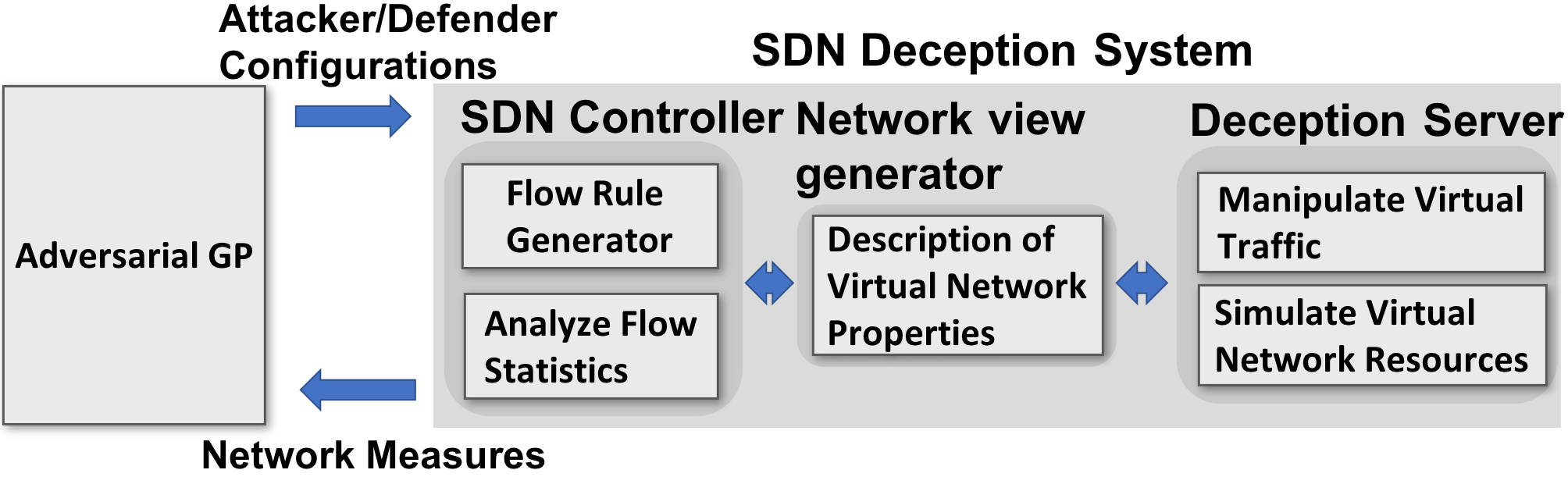}
  \caption{\DarkHorse fitness evaluation(left side) has a defensive
  SDN system based on the POX SDN Controller that creates different
  network views and manipulates traffic, and an attacker that performs
  NMAP scans in a \texttt{mininet} simulation. The scan results are passed into
  a fitness function that assigns fitness values for the adversaries'
  configurations. The search in the \compCA uses Maximum Expected
  Utility of the detection time to assign a fitness value.}
\label{fig:dark_horse}
\end{figure}

Once an adversary has compromised a network endpoint, they can perform
network reconnaissance~\cite{sood:targeted}. After reconnaissance
provides a view of the network and an understanding of where
vulnerable nodes are located, attackers are able to execute a plan of
attack. One way to protect against reconnaissance is by obfuscating
the network to delay the attacker. This approach is well suited to
software defined networks (SDN) such as those deployed in cloud
server settings because it requires programmability that they
support~\cite{kirkpatrick:networking}.  The SDN controller knows which
machines are actually on the network and can superficially alter
(without function loss) the network view of each node, as well as
place decoys (honeypots) on the network to mislead, trap and slow down
reconnaissance. \DarkHorse is shown in Fig.~\ref{fig:dark_horse}.

One such multi-component deceptive defense system
\cite{achleitner:cyber} foils scanning by generating ``camouflaged''
versions of the actual network and providing them to hosts when they
renew their DHCP leases.  We use this deception system and
\texttt{mininet}~\cite{mini:mininet} within the framework as an engagement
environment. This allows us to explore the dynamics between attacker
and defender on a network where the deception and reconnaissance
strategies can be adapted in response to each
other~\cite{pertierrathesis2018}.

A deception strategy is executed through a modified \texttt{POX} SDN
controller. A reconnaissance strategy is executed by an NMAP
scan\cite{Lyon:nmap}. The attacker strategy includes choices of: which
IP addresses to scan, how many IP addresses to scan, which subnets to
scan, the percent of the subnets to scan, the scanning speed, and the
type of scan.  The defender strategy includes choices of: the number
of subnets to set up, the number of honeypots, the distribution of the
real hosts throughout the subnets, and the number of real hosts that
exist on the network. Fitness is comprised of four components: how
fast the defender detects that there is a scan taking place, the total
time it takes to run the scan, the number of times that the defender
detects the scanner, and the number of real hosts that the scanner
discovers. Through experimentation and analysis, the framework is able
to discover configurations that the defender can use to
significantly increase its ability to detect scans. Similarly, there
are specific reconnaissance configurations that have a better chance
of being undetected.

\subsection{\RIVALS Summary}

We summarize \RIVALS use cases with Table~\ref{tab:rivals_framework} and its architecture with Fig.~\ref{fig:rivals_concepts}. The \RIVALS framework is one example of \TOPIC, where the domain of network security is one of many possibilities in \cybersecurity. Possible future steps are elaborated in the next section.

\begin{table}
  \centering
  \caption{How \RIVALS components are configured to express specific use~cases.}
  \label{tab:rivals_framework}
  \begin{tabular}{l|p{2cm}|p{2.7cm}|p{2.7cm}}
  \toprule
     & \multicolumn{3}{c}{\textbf{\textit{Use~Case}}} \\
    \cmidrule{2-4} 
    \textbf{Component} & \textbf{Robustness vs Denial} & \textbf{Deception vs Reconnaissance} & \textbf{Isolation vs Contagion} \\
    \textit{NAME} & \SCATTERED & \DarkHorse & \AVAIL \\
    \midrule
    Threat Environment & P2P vs DDOS & Pox SDN vs Nmap & Enclave vs Malware \\
    Evaluation & ALFA sim & Mininet & High level Sim \\    
    Objective & Mission disruption & Detection speed & Mission delay \\
    Behavior: Attacker & Node/Edge impairment & Scanning parameters & Strength \& Duration \\
    Behavior: Defender & Network settings & Honeypots & Network taps \& device placement \\
    Adaptivity & Yes & No & No \\
    Text description & Section~\ref{sec:dos-attacks-peer} & Section~\ref{sec:intern-reconn-softw} & Section~\ref{sec:avail-attacks-segm} \\
  \bottomrule
  \end{tabular}
\end{table}

\section{Taking Stock and Moving Forward}
\label{future:past}

\paragraph{Summary}

Considering the paper from bottom to top, we described three use cases: \SCATTERED, \AVAIL, and \DarkHorse. For 3 specific network security contexts, they investigate unique threat environments, 
objectives and behaviors for their adversary species, and varying capacities to adapt during an attack on a mission. They draw upon the \RIVALS framework
in which adversarial arms races are the activity of interest. \RIVALS is an example of \textit{\TOPIC} 
and is motivated by a call to arms to work on automated defensive reaction to waves of attacks.  \RIVALS pushes towards this goal by considering how \textit{both attack and defense} adapt to each other.  While we identified a breadth of applications to cyber security in the set of EAs we surveyed, we found just a modest number of extant \textit{adversarial GP} that represented approaches for behavior investigation and two population dynamics, the latter combining competitive coevolution with GP.  The body of prior work relevant to  cyber security from AI and games and software engineering more generally informs the reader as to how \compCA{s} and EAs can be used as building blocks in different adversarial contexts. Finally, at the top, we argue that \textit{\TOPIC} is compellingly worthy of future community attention because of its potential to solve a critical and growing set of challenges in cyber security and because of how well GP and adversarial evolution match the technical nature of the challenges.

\paragraph{Future directions}

To end, we consider what future research questions and directions of study, stemming from the domain of cyber security, are prompted by \textit{Adversarial GP}.  

We forecast explorations driven by the wide array of cyber security scenarios will arise.  There is a plethora of attack surfaces. We have mentioned some in
Section~\ref{sec:security}, however, there are others such as networked cyber physical
systems, ransomware, gadgets and malware.  For example, how can GP help examine the adversarial cyber security of a power grid or self-driving cars? How could ransomware's next mutation be predictable? Can GP leverage code repositories and gadgets to discover new malware?  Social engineering is currently an Achilles heel for security measures because it preys on human nature and once privileged access has been obtained, it is much harder to discern an attacker. Insider threats similarly disguise attacks. How can GP inform insider threats and social engineering?  

One challenge for the GP community is to become adequately conversant in cyber security topics (though no member needs to be conversant in all of them). Collaborations with cyber security researchers seem advisable, offering additive and synergistic benefits. It will be important to identify the most appropriate level of abstractions from which to design GP languages or function and terminal sets. It will be imperative to develop metrics of success and paths to technology transfer and solution deployment.  At a practical level, undoubtedly new modeling and simulation modeling platforms will be needed. They will need to be general purpose (somewhat like \RIVALS) and specific.  Undoubtedly some will need to be scaled due to the computational cost of executing a mission.

We forecast the opportunity for extensions and innovations in GP and coevolutionary algorithms for cyber security.  Different patterns and rates of evolution will be observed across different security scenarios.  The opening to develop new algorithmic models of these phenomena is exciting. GP evolves behavior but many different expressions of behavior remain unexplored. Developing the new behavioral evolutionary capabilities that will be needed is both challenging and motivating.

We also forecast that research into \textit{\TOPIC} will push bridge building to other research areas. One impetus will be the quest to describe at a finer granularity what is happening during cyber evolution.  For this, it may be advisable to consider evolution through the lens of individuals, such as how Artificial Life (ALife)~\cite{macal2010tutorial} models individuals. What cyber security scenario should be studied with ALife? How can agents in a cyber security ALife system be represented by evolvable GP behaviors? What if a GP defender ``died'' after an attack and attackers starve as they fail to penetrate defenses? How would dynamics at this scale offer different insights?  What would a simulation of the DDOS attack ecosystem reveal  about the progression of botnet sizes if intra-species competition was examined? What can ALife studies into ecosystem dynamics around intra-species competition and cooperation reveal? 

Arguably, adversarial GP is a form of agent-based modeling (ABM,~\cite{moran2017effects}) where agents are GP executable structures. Some ABM systems model more complex domains while the agents have simpler strategy spaces. Are there ABM investigations such as those of financial markets, crowd control, infectious diseases or traffic simulation that suggest innovative transfer to Adversarial GP and cyber security?  Both ALife and ABM have examples that consider the spatial dimension of a domain. \CompCA{s}  have spatial models. Interesting new models lie at these intersections.

Another area with potentially valuable interaction is Machine Learning (ML).  Statistical machine learning relies on  retrospective training
data to learn a model capable of generalizing to new unseen data drawn
from the same distribution as the retrospective data. In contrast, \textit{\TOPIC} is not data driven and this conveys it a niche. However, what can be transferred from the studies that
have started to bridge supervised learning and test-based
co-optimization\cite{popovici2017bridging}?
Adversarial classification ~\cite{dalvi2004adversarial} as a game
between classifier and adversary also appears in ML. 
As well, Generative Adversarial Networks~(GANs)~\cite{goodfellow2014explaining}
 derive generative models from samples of a distribution using
adversarial neural network training.  Coevolution has been applied to improve the
robustness and scalability of these
systems~\cite{schmiedlechner2018towards}. Competitive coevolution has also been combined with
multi-layer perceptron classifiers with floating point
representation~\cite{castellani2018competitive}.

Studies of adversarial coevolution in cyber security based on data that
classify attacks using machine learning methods exist, e.g. see a pro-active defense for evolving cyber threats and
a predictive defense against evolving adversaries in~\cite{colbaugh2011proactive,colbaugh2012predictive,colbaugh2013moving}.
In adversarial Machine Learning the attacker manipulates data in order
to defeat the Machine Learning  model~\cite{lowd2005adversarial,huang2011adversarial,biggio2017wild}. Can coevolution be used to anticipate continual adversarial evasion and guide model hardening?
 
Game theory considers equilibria and paths to equilibria.  What is an equilibrium in cyber security? It could be the complete wipeout of one side of the adversarial equation. Or, it could be the point where an attacker chooses not to target a defense because it is less expensive to go elsewhere. It could be a Nash equilibrium where neither attacker or defender has a better tactical position to go to unless the other simultaneously changes also. One direction should investigate how these game theoretic notions can inform the highly empirical, algorithmic systems of Adversarial GP.

We believe the future directions are not enumerable because the likelihood of new adaptations by adversaries will never be zero.  To date, defenses expose far more attack surface than they can protect and attackers only need to find one crack to penetrate. To this end, the emerging importance of \textit{\TOPIC} is not likely to fade. We trust this justifies our call to arms.

%

\section*{Acknowledgements}
This was supported by the CSAIL CyberSecurity Initiative. This material is based upon work supported by DARPA.  The views and conclusions contained herein are those of the authors and should not be interpreted as necessarily representing the official policies or endorsements. Either expressed or implied of Applied Communication Services, or the US Government. 
This project has received funding from the European Union’s Horizon 2020 research and innovation programme under the Marie Skłodowska-Curie grant agreement No 799078.

\bibliographystyle{spmpsci}      
\bibliography{refs,GPEM_extra_bibliography}   

\begin{appendices}

\section{}

\subsection{Firefly squid bioluminescence defense}
\label{app:firefly-squid}
When firefly squid are seen from below by a predator, the bioluminescence helps to match the squid's brightness and color to the sea surface above.

\subsection{A coevolutionary perspective of the U.S. automotive industry}
\label{app:a}

Talay, Calantone, and Voorhees presented an study that explicitly terms competitive interactions between firms ``Red Queen competition'', in which gains from innovations are relative and impermanent~\cite{doi:10.1111/jpim.12080}.

\subsection{Advanced Persistent Threats and Ransomware}
\label{app:apts-ransmoware}

Some DOS attacks includes Advanced Persistent Threats (APT) and Ransomware.
The first ones have multiple stages starting at external reconnaissance then moving to intrusion (e.g. social engineering or use of zero day exploits), laterally moving malware, command and control direction to data exfiltration and, finally, self-erasure. Ransomware, which largely preys upon unpatched systems and which exploits anonymous payment channels enabled by Bitcoin, has also recently become more frequent.

\subsection{Cyber security attack categorization}
\label{app:cyber-secur-attack}
Examples of attacks, classifications and taxonomies can be found at
\url{https://cwe.mitre.org/index.html}. One categorization is:
\begin{inparaenum}[\itshape A)]
\item Advanced Persistent Threats, ``lurking'' threats from
  resourceful persevering adversaries
\item Denial of Service Attack, defense resource limitation and exposure,
  means of penetrating systems
\item Identity theft, e.g. impersonating users.
\end{inparaenum}
Attacks are also characterized by their stages on a timeline. Another
characterization is based on the attacker identity, from individuals
to organized criminals and nation states, and what resources they
access, see ~\cite{bodeau2013characterizing} for details.

\end{appendices}

\end{document}